\documentclass{aa}

\usepackage[varg]{txfonts}
\usepackage{graphicx}
\usepackage{natbib}
\bibpunct{(}{)}{;}{a}{}{,}

\newcommand{\msun}{\mbox{$\rm M_\odot$}}

\newcommand{\gammaray}{\mbox{$\gamma$-ray}}

\begin{document}

\title{Cosmic rays and thermal instability in self-regulating cooling flows of massive galaxy clusters}

\titlerunning{Cosmic rays in galaxy clusters}
\authorrunning{Beckmann et al.}

\author{Ricarda S. Beckmann\inst{1,2}\thanks{ricarda.beckmann@iap.fr} \and Yohan Dubois\inst{1} \and Alisson Pellissier \inst{3}  \and Valeria Olivares \inst{4,5} \and Fiorella L. Polles \inst{4,6}  \and Oliver Hahn\inst{7,8} \and Pierre Guillard \inst{1,9} \and Matthew D. Lehnert \inst{10} }
\institute{Institut d'Astrophysique de Paris, CNRS, Sorbonne Universit\'{e}, UMR7095, 98bis bd Arago, 75014 Paris, France
\and Institute of Astronomy and Kavli Institute for Cosmology, University of Cambridge, Madingley Road, Cambridge CB3 0HA, UK
\and Laboratoire Lagrange, Observatoire de la Côte d’Azur, CNRS, Université Côte d’Azur, UMR7293, Boulevard de l’Observatoire, CS 34229, 06304 Nice, France
\and LERMA, Observatoire de Paris, PSL Research Univ., CNRS, Sorbonne Universit\'e, UMR8112, 75014 Paris, France
\and Department of Physics and Astronomy, University of Kentucky, 505 Rose Street, Lexington, KY 40506, USA
\and SOFIA Science Center, USRA, NASA Ames Research Center, M.S. N232-12 Moffett Field, CA 94035, USA
\and Department of Astrophysics, University of Vienna, Türkenschanzstrasse 17, 1180 Vienna, Austria
\and Department of Mathematics, University of Vienna, Oskar-Morgenstern-Platz 1, 1090 Vienna, Austria
\and
Institut Universitaire de France, Ministère de l'Education Nationale, de l'Enseignement Supérieur et de la Recherche, 1 rue Descartes, 75231 Paris Cedex 05, France
\and Centre de Recherche Astrophysique de Lyon, ENS de Lyon, Université Lyon 1, CNRS, UMR5574,  69230 Saint-Genis-Laval, France}

\date{Accepted . Received ; in original form }

\abstract{One of the key physical processes that helps prevent strong cooling flows in galaxy clusters is the continued energy input from the central active galactic nucleus (AGN) of the cluster. However, it remains unclear how this energy is thermalised so that it can effectively prevent global thermal instability. One possible option is that a fraction of the AGN energy is converted into cosmic rays (CRs), which provide non-thermal pressure support, and can retain energy even as thermal energy is radiated away. By means of magneto-hydrodynamical simulations, we investigate how CR injected by the AGN jet influence cooling flows of a massive galaxy cluster. We conclude that converting a fraction of the AGN luminosity as low as 10\% into CR energy prevents cooling flows on timescales of billion years, without significant changes in the structure of the multi-phase intra-cluster medium. CR-dominated jets, by contrast, lead to the formation of an extended, warm central nebula that is supported by CR pressure. We report that the presence of CRs is not able to suppress the onset of thermal instability in massive galaxy clusters, but CR-dominated jets do significantly change the continued evolution of gas as it continues to cool from isobaric to isochoric. The CR redistribution in the cluster is dominated by advection rather than diffusion or streaming, but the heating by CR streaming helps maintain gas in the hot and warm phase. Observationally, self-regulating, CR-dominated jets produce a \gammaray~ flux in excess of current observational limits, but low CR fractions in the jet are not ruled out.} 


\maketitle



\section{Introduction}

From X-ray observations, the hot gas that fills galaxy clusters, the so-called intra-cluster medium (ICM), is cooling rapidly and is expected to produce a cooling flow of about 100 - 1000 $\msun \,\rm yr^{-1}$~\citep{Fabian1994}. However, observed star formation rates (SFRs) in clusters reach only about 1-10\% of the inferred cooling rates \citep{McDonald2018}. Clusters must therefore contain a heating source that prevents over-cooling and slows star formation down while still allowing for the formation of the observed warm H$\alpha$ emission nebulae, as well as cold molecular gas and their extended filamentary morphologies~\citep[e.g.][]{Salome2011,Olivares2019, Olivares2022}. Research into the contribution of different heating mechanisms, from feedback driven by active galactic nuclei (AGN) \citep[e.g.][]{Sijacki2006,Dubois2010,Gaspari2011,Li2015,Voit2017,Beckmann2019b,Ehlert2022} via thermal conduction \citep[e.g.][]{Bogdanovic2009,Parrish2009,Ruszkowski2011,Kannan2017} and turbulence \citep[e.g.][]{Kunz2011,Voit2018} to cosmic rays \citep[CRs;][e.g.]{Sijacki2008,Ruszkowski2017,Ehlert2018}, has shown that long-term self-regulation of cooling flows is a complex thermodynamic problem that most likely relies on the interplay between different mechanisms that are active at the same time.  

Cosmic rays in particular have the potential to significantly change the properties of the multi-phase gas in the cluster center, as they  provide non-thermal pressure support and can retain energy even as thermal energy is radiated away. High-energy CRs are accelerated in strong shocks~\citep[e.g.][]{Bell1978} that occur in and around AGN jets \citep{Blandford1978,Blandford2000,Matthews2019}.

On large scales, it has been suggested that variations in CR transport physics are responsible for the observed dichotomy of galaxy cluster profiles \citep{Guo2008,Ensslin2011}, which appear as either cool-core or non-cool-core clusters. CR have the potential to offset cooling flows by providing a source of stable heating \citep{Fujita2011,Jacob2017a}, but these steady-state solutions require a CR pressure beyond current upper observational limits for clusters with large cooling radii or high SFR \citep{Jacob2017b}.

On small scales, \citet{Ruszkowski2018} showed that CR could provide the missing heating source that maintains observed H$\alpha$ filaments in galaxy clusters at their observed temperature of $10^4\,\rm K$. Using linear stability analysis, \citet{Kempski2019} showed that the presence of CRs modifies local thermal instability criteria, while \citet{Butsky2018} studied how the presence of CRs influences the morphology of condensed gas. \citet{Huang2022} showed that the presence of CRs reduces the density contrast of thermally unstable gas.

Early simulations of isolated galaxy clusters by \citet{Sijacki2008} showed that CRs injected by AGN, even in the absence of CR transport and heating mechanisms beyond advection, allow clusters to have cool cores without strong cooling flows merely by providing non-thermal pressure support. More complete recent simulations that include CR transport processes (including diffusion and streaming), CR heating, and a variety of different CR injection mechanisms from the AGN jet \citep{Ruszkowski2017,Wang2020,Su2021} to volume-filling injection \citep{Su2019}, showed that CRs with streaming-instability heating improve cluster self-regulation and allow for the formation of cold gas without a run-away cooling catastrophe. \citet{Su2021} showed that a large amount of CR injection leads to over-quenching, in which cluster cooling is completely suppressed over long periods of time, in contrast to observations. \citet{Ji2020} used cosmological simulations to show that the impact of CRs is not only limited to the cluster centre, but also produces warm volume filling gas at large radii from the centre. 

One piece of observational evidence for CR playing a part in galaxy cluster cooling cycles comes from the observations of Faranoff-Riley I (FRI) jet lobes. To be in pressure equilibrium with the surrounding ICM, FRI jet lobes need a dominant non-thermal pressure component  \citep{Morganti1988,Worrall2009,Croston2008}, with CRs being the most likely candidate \citep{Croston2018}. Another way to observationally search for CRs is via the \gammaray~ emission produced by the decay of pions. So far, only upper limits have been detected for such pionic $\gamma$ rays from the ICM of galaxy clusters. In the Coma cluster, combined \gammaray~ and radio observations place the maximum ratio of CR pressure to thermal pressure, $\eta = P_{\rm CR} / P_{\rm therm} <0.1 $ for both hadronic and reacceleration models \citep{Ackermann2010,Aleksic2012,Brunetti2012,Ackermann2014,Ackermann2016,Brunetti2017}. For the Perseus cluster, constraints are even tighter, with $\eta < 0.02$ if the CR density profile is the same as the gas pressure profile. However, if the CR density profile is overall flatter than the gas pressure profile, that is, if CRs preferentially escape from the cluster centre,  $\eta < 0.2 $ is supported by the data \citep{Ahnen2016}. Finally, CRs can be detected in galaxy clusters via the Sunyaev-Zeldovich effect \citep{Abdulla2019,Ehlert2019}.

We investigate how CRs injected by the AGN jet in the cluster centre influence the onset of thermal instability in the multi-phase ICM and the cooling flow of a massive galaxy cluster. The paper is structured as follows: A theoretical discussion of the expected impact of CRs on cluster cooling flows is presented in Section \ref{sec:theory}, while a comparative analysis using simulations (parameters shown in Section \ref{sec:simulation}) is presented in Section \ref{sec:suite}. The impact of CRs on the multi-phase gas and local thermal instability is shown in Section \ref{sec:phase}. An observational comparison is presented in Section \ref{sec:observation}, followed by  our conclusions (Section \ref{sec:conclusions}). The robustness of the results with respect to small perturbations is investigated in Appendix \ref{sec:stochastic}.

\section{Thermal instability in clusters}
\label{sec:theory}
\subsection{Review}

To avoid massive cooling flows, clusters need to be in global hydrostatic equilibrium. However, clusters are not described by perfectly uniform density profiles, but instead are turbulent, with local variations in density, temperature, and velocity. This leads to pockets of local thermal instability in which the hot ICM is able to cool and condense. 

In the absence of magnetic fields, gas becomes locally unstable when the local cooling time
\begin{equation}
    t_{\rm cool} = \frac{2}{3} \frac{n k_{\rm B} T}{n_{\rm e} n_{\rm i} \Lambda}
    \label{eq:tcool}
\end{equation}
is shorter than the free-fall time
\begin{equation}
    t_{\rm ff}=\sqrt{\frac{2r}{g}}\, , 
    \label{eq:tff}
\end{equation}
where $n$ is the gas number density, $n_{\rm e}$ and $n_{\rm i}$ are the electron and ion number density, respectively, $k_{\rm B}$ is the Boltzmann constant, $T$ is the gas temperature, $\Lambda$ is the cooling function, $g$ is the local gravitational acceleration, and $r$ is the radius from the cluster centre. 

Using linear perturbation theory, \citet{McCourt2012} showed that local thermal instabilities occur when $f_{\rm ff} = t_{\rm cool}/t_{\rm ff} \leq 1 $ if there is a  global heating function that ensures global thermal equilibrium. Simulations have generally found that $f_{\rm ff}$ can be as high as $f_{\rm ff} = 10-25$ for spherically averaged profiles 
 \citep{Sharma2012,McCourt2012,Beckmann2019b}. This is due to the dispersion of entropy perturbations \citep{Voit2021}, which lead to a globally elevated $t_{\rm ff}$ despite the fact that  $f_{\rm ff}=1$ locally~\citep{PalChoudhury2019,Voit2021}. Observationally, it has been confirmed that dense gas is located at the minima of the radial profile of $f_{\rm ff}$ \citep{Hogan2017,Pulido2018,Olivares2019,Olivares2022}.

Another approach to determining local thermal instability is to follow \citet{Gaspari2018} and consider the local eddy turnover time, 
\begin{equation}
    t_{\rm eddy}=2 \pi \frac{r^{2/3}L^{1/3}}{\sigma_v}
,\end{equation}
in comparison to the cooling time, where $\sigma_v$ is the velocity dispersion at the injection scale $L$. This criterion is based on the assumption that the kinematics of the warm ionized gas correlates linearly with those of the hot medium, following the development of a multi-phase cascade of turbulent energy. However, there have recently been insights from observations \citep{Li2020} and simulations \citep{Wang2021,Mohapatra2022} that this assumption might not hold in galaxy clusters.

In the presence of non-thermal energies beyond turbulence, such as when magnetic fields or CR are taken into account, the thermal instability changes as the non-thermal energies can provide additional pressure support against collapse. CR energy also is not as susceptible to efficient radiative cooling as thermal energy. 

Studying the impact of magnetic fields on thermal instability, \citet{Ji2018} used idealised numerical experiments similar to those used in \citet{McCourt2012} to quantify the impact of magnetic fields on thermal instability. Rather than define a clear threshold for $f_{\rm ff}$, \citet{Ji2018} took a more continuous approach and measured the magnitude of the density fluctuations, $\delta \rho / \rho$ as a function of $f_{\rm ff}$. For the hydro-only case, they reported
\begin{equation}
    \frac{\delta \rho }{ \rho} \Bigg|_{\rm hydro} = 0.1 f_{\rm ff} ^{-1}
.\end{equation}
In the presence of magnetic fields, this becomes 
\begin{equation}
    \frac{\delta \rho }{ \rho} \Bigg|_{\rm MHD} = 3 \beta^{-1/2} f_{\rm ff} ^{-1}
,\end{equation}
where $\beta= {8 \pi P_{\rm therm} }/{B^2}$ is the plasma beta and $P_{\rm therm}$ is the thermal pressure. Comparison of the two shows that $\frac{\delta \rho }{ \rho} \big|_{\rm MHD} > \frac{\delta \rho }{ \rho} \big|_{\rm hydro}$ for fields as weak as $\beta \leq 1000$. Therefore even weak magnetic fields produce stronger density fluctuations in the gas. 

Adding CRs is somewhat more complicated, as they bring a range of energy transfer mechanisms that have competing effects on thermal instability. \citet{Kempski2019} conducted a linear stability analysis of a plasma in the presence of CRs and concluded that the unstable regime depends on the relative importance of different CR transport processes, such as diffusion and streaming, and on the importance of CR heating. Results depend on three parameters:
\begin{enumerate}
\item the slope of the cooling function $\mathcal{L}_{\rm th}$ with respect to temperature $T$,
\begin{equation}
   \alpha_{T} =  \frac{\partial \log \mathcal{L}_{\rm th}}{\partial \log T}  \,
,\end{equation}
which for the simulations presented here is shown in Appendix \ref{sec:cooling_function};
\item the ratio of CR pressure $P_{\rm CR}$ and thermal pressure $P_{\rm therm}$:
\begin{equation}
    \eta = \frac{P_{\rm CR}}{P_{\rm therm}}\, ;
\end{equation}
\item and the ratio
\begin{equation}
    \chi=\frac{D_{\rm CR} }{ t_{\rm cool} \eta u_{\rm A}^2}\, ,
\end{equation}
where $D_{\rm CR}$ is the CR diffusivity, and $u_{\rm A}$ is the Alfv\'en velocity, $\chi$, which measures the relative strength of the cosmic ray heating rate to the radiative thermal cooling rate, encapsulates the complex interplay between CR diffusion, cooling, and the local energy distribution, which all play a part in  determining the local thermal (in)stability of gas in the presence of CRs.
\end{enumerate}

In general, the role of CRs in thermal instability depends on the value of $\chi$. When $\chi > 1$, thermal instabilities develop with an amplitude proportional to $2-\alpha_{T}$, that is, thermal instabilities grow strongly when $\alpha_{T}<2$. In the opposite regime, where $\chi\leq1$, CR diffusion suppresses thermal instability below a maximum length-scale, which can be expressed as an effective field length of the form
\begin{equation}
\lambda_{\rm CRF} = \left\{
    \begin{array}{ll}
         2\pi |\vec{b}. \vec{\tilde{k}}| \sqrt{\eta \rm D_{\rm CR} t_{\rm cool}} & \eta\le 1  \, , \\
         2\pi |\vec{b}. \vec{\tilde{k}}| \sqrt{\frac{\rm D_{\rm CR} t_{\rm cool}}{\eta}} & \eta>1  \, ,\\
    \end{array}\right.
    \label{eq:CR_field_length}
\end{equation}
where $\vec{b}$ is the magnetic field unit vector, and $\vec{\tilde{k}}$ is the wave number unit vector. The thermal instability continues to occur on larger scales. $\lambda_{\rm CRF}$ is at a maximum when $\eta=1$, and it falls off for higher and lower values.

\subsection{Expected behaviour for galaxy clusters}

In this section, we predict $\eta_{\rm crit}$, the value of $\eta$ required to have $\chi=1$ for galaxy clusters of different masses in order to discuss the potential impact of CRs on thermal instability within different regions of the cluster. In galaxy clusters, temperature, density, and magnetic field strength vary strongly with radius, as does CR pressure. As a consequence, the $\eta$ required to have $\chi=1$ is expected to be a function of radius. In turn, this means that the impact of CRs on thermal instability will vary throughout the cluster. We would expect CRs to be unable to suppress thermal instability in massive galaxy clusters if $\chi>1$, as hot gas in galaxy clusters cools predominantly through free-free emission, which has $\alpha_{T} = 0.5 < 2$. By contrast, if $\chi<1$ for a significant volume of the cluster, there could be a maximum length scale, the CR Field length $\lambda_{\rm CRF}$ (see Eq. \ref{eq:CR_field_length}), below which CRs are able to suppress thermal instability. Instability would be expected to continue on scales larger than $\lambda_{\rm CRF}$. As both $\chi$ and $\lambda_{\rm CRF}$  explicitly depend on $\eta$, the impact of CRs on thermal instability in galaxy clusters depends very strongly on the radial distribution of CRs, as expected.

\begin{figure}
    \centering
    \includegraphics[width=\columnwidth]{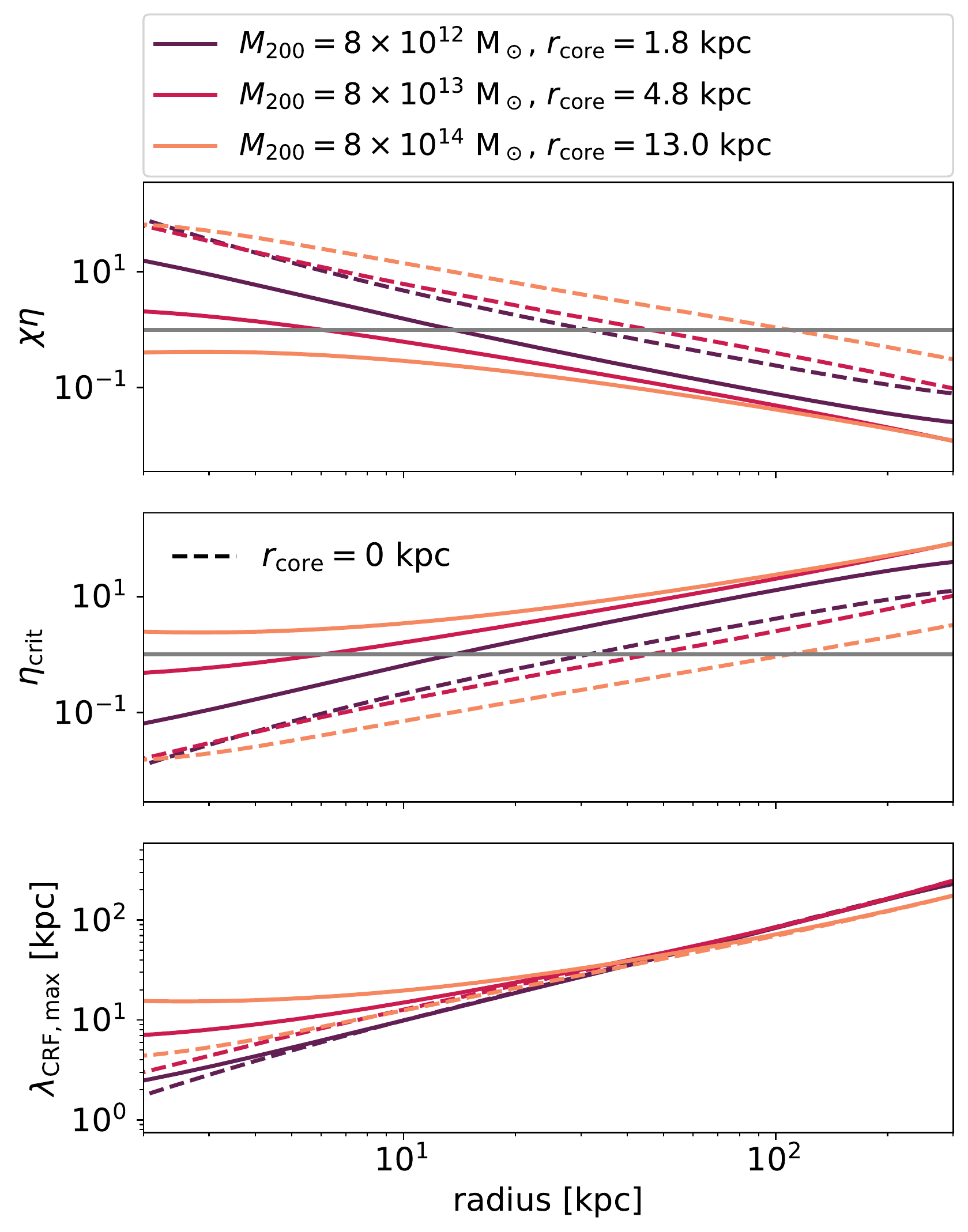}
    \caption{Radial profile of $\chi\eta$ (top panel), $\eta_{\rm crit}$, the minimum $\eta$ required for the strong diffusion regime (middle panel), and the maximum $\lambda_{\rm CRF}$ (bottom panel) for the galaxy clusters of different masses in hydrostatic equilibrium. Different colours stand for different halo masses as indicated in the top inset, either with a cored (solid lines) or with a non-cored profile (dashed lines). Grey horizontal lines demarcate the two regimes in $\chi$ (top panel) and gas that are CR-pressure dominated vs thermal pressure dominated (middle panel). $\eta_{\rm crit}$, the minimum value required for $\chi<1$ increases with radius. By contrast, CR pressure profiles are expected to decrease with radius. It is therefore unlikely that $\chi<1$ at large radii, which means that CRs are unable to stabilise small-scale instabilities outside the cluster core.}
    \label{fig:analytic_chi_eta}
\end{figure}

In Fig. \ref{fig:analytic_chi_eta} we show example profiles of $\chi\eta$ for galaxy clusters of different masses in hydrostatic equilibrium. To test the impact of different distributions of CRs on the cluster, we computed  $\chi\eta$ and then determined $\eta_{\rm crit}$ , which will give $\chi<1$ if $\eta > \eta_{\rm crit}$ (or $\chi >1 $ if $\eta < \eta_{\rm crit}$). Profiles to compute $\chi\eta$ were computed in the same way as the initial conditions for the simulations presented in this paper (see Section \ref{sec:initial_conditions}). Like for the initial conditions, density profiles are based on a cored Navarro-Frenk-White (NFW) profile \citep{Navarro1997}. The concentration parameter, gas fraction, and black hole mass are scaled with cluster mass according to \citet{Maccio2007}, \citet{Andreon2017}, and \citet{Bogdan2017}, respectively. The central magnetic field was set to $10 \, \mu \rm G$~\citep[e.g.][]{GOVONI2004}, and was scaled with $\rho(r)^{2/3}$. For each cluster, a cored (solid) and a non-cored (dashed) profile is shown. The solid orange lines in Fig. \ref{fig:analytic_chi_eta} correspond to the profiles we used as initial conditions.

Fig. \ref{fig:analytic_chi_eta} shows that the profile of $\chi\eta$ varies strongly with galaxy cluster mass, thermal profile (cored or non-cored), and radius. More massive clusters need higher CR pressure to reach $\chi<1$, which requires $\eta > \eta_{\rm crit}$ (middle panel) as $\chi \propto \frac{1}{\eta}$. The required value of $\eta_{\rm crit}$ increases with radius for all cluster masses. 

One source of CR, injection via supernovae and AGN jets, is concentrated in the cluster centre and would cause $\eta$ to decrease with radius. From such injections, it is likely that only the cluster centre, or no part of the cluster, has $\chi < 1,$ where CRs can potentially stabilise against thermal instability. Another potential source of CR injection are large-scale structure formation shocks on megaparsec scales \citep{Ryu2003,Pfrommer2006,Vazza2009}. However, energy ratios of CR energy and thermal energy for these circum-cluster shocks are low because strong shocks are rare in and around galaxy clusters \citep{Vazza2009}, and the injection scales are very far from the cluster centre. It is therefore unlikely that these CRs make a significant difference to $\eta$ in the cluster centre. Outside regions where $\chi <1$, even in the presence of CRs, thermally unstable gas will remain, so that even in the presence of CRs as $\alpha_{T} = 0.5 < 2$ for the free-free emission that is the dominant cooling channel for hot cluster gas.

Inside this potentially CR-stabilised core, CRs can suppress thermal instability on scales below  $\lambda_{\rm CRF}$ (see Eq. \ref{eq:CR_field_length}). The bottom panel of Fig. \ref{fig:analytic_chi_eta} shows that the maximum length scale below which CRs can suppress thermal instability in a massive cluster is about 10 kpc or more, which occurs if $\eta = 1$. $\lambda_{\rm CRF}$ decreases for higher and lower values of $\eta$, and for dimensions that are not parallel with the magnetic field lines. For a Perseus-like cluster, 10 kpc of $\lambda_{\rm CRF}$ is comparable to the observed length of filaments, but much larger than their 10 pc width \citep{Conselice2001,Fabian2016}. This lends credibility to the idea that observed filaments might be aligned with magnetic field lines. 

As $\eta_{\rm crit}> 1$ for massive galaxy clusters at all radii, there will be gas that is dominated by CR pressure, but in which CRs will be unable to suppress thermal instability, that is, gas for which $1<\eta <\eta_{\rm crit}$. However, CRs influence not only the local thermal instability, but also the thermodynamic evolution of the gas after the onset of thermal instability.  Gas with $\eta>1$ cools isochorically rather than isobarically. As gas cools, $P_{\rm therm}$ decreases, thus, $\eta = {P_{\rm CR}}/{P_{\rm therm}}$ increases even if $P_{\rm CR}$ is constant. 

Overall, we conclude that CRs might be able to suppress thermal instability on relevant length scales in the cluster core, but are unlikely to be able to do so outside this central region even for gas dominated by CR pressure. Understanding where and how CRs are able to change the thermal instability criterion in massive galaxy clusters, and how the evolution of this newly condensed gas differs in the presence of CRs from the CR-free case is the aim of this paper.

\section{Simulation}
\label{sec:simulation}

\subsection{Magneto-hydrodynamics with cosmic rays}
\label{sec:CRphysics}

This paper presents a set of hydrodynamical and magneto-hydrodynamical (MHD) simulations of isolated galaxy clusters with and without CRs. The cluster simulations were run with the adaptive mesh refinement code {\sc ramses}~\citep{Teyssier2002} that solves for the MHD~\citep{Fromang2006} with CRs and two ion-electron temperatures~\citep{Dubois2016},
\begin{align}
    &\frac{\partial \rho}{\partial t} + \nabla . (\rho \vec{u}) = 0\,,\\
    &\frac{\partial \rho \vec{u}}{\partial t} +\nabla \left( \rho \vec{u}\vec{u}+P_{\rm tot} + \frac{\vec{B} \vec{B}}{4 \pi}\right)=0\,,\\
\label{eq:total}
    &\frac{\partial e}{\partial t} +\nabla. \left( (e+P_{\rm tot}) \vec{u} - \frac{\vec{B} (\vec{B}. \vec{u})}{4 \pi}\right) =  \mathcal{L}_{\rm rad}  \nonumber \\
    &\hspace{0.8cm}  - \nabla . \vec{F}_{\rm st} -\nabla. \vec{F}_{\rm d} -\nabla. \vec{F}_{\rm cond}\,, \\
    & \frac{\partial \vec{B}}{\partial t} - \nabla \times (\vec{u}\times\vec{B}) = \vec{0}\,, \\
\label{eq:CR}
    &\frac{\partial e_{\rm CR}}{\partial t} +\nabla.( e_{\rm CR} \vec{u}) = -P_{\rm CR}\nabla. \vec{u}  + \mathcal{L}_{\rm CR} \nonumber \\ 
    & \hspace{0.8cm} - \nabla . \vec{F}_{\rm st} -\nabla. \vec{F}_{\rm d} -\vec{u}_{\rm{st}} . \nabla P_{\rm CR}\, , \\
\label{eq:electron}
    &\frac{\partial e_{\rm e}}{\partial t} +\nabla.( e_{\rm e} \vec{u}) = -P_{\rm e}\nabla. \vec{u} -\nabla. \vec{F}_{\rm cond} + Q_{\rm e\leftrightarrow i} + \mathcal{L}_{\rm th}\, , 
\end{align}
where $\rho$ is the mass density; $\vec u$ is the velocity; $e=0.5\rho \vec{u}^2+e_{\rm e}+e_{\rm i}+e_{\rm CR}+\vec{B}^2/8\pi$ is the total energy density; $e_{\rm e}$, $e_{\rm i}$ , and $e_{\rm CR}$ are the electron, ion, and CR energy densities, respectively; $P_{\rm tot}=P_{\rm e}+P_{\rm i}+P_{\rm CR}$ is the total gas pressure; $P_{\rm e}$, $P_{\rm i}$ , and $P_{\rm CR}$ are the electron, ion, and CR pressures, respectively; and $\vec B$ is the magnetic field. Each pressure component is related to its corresponding energy density assuming an adiabatic equation of state with adiabatic index $\gamma_{\rm e}=\gamma_{\rm i}=\gamma=5/3$ and $\gamma_{\rm CR}=4/3$.

In the CR energy equation (eq.~\ref{eq:CR}), $\vec{F}_{\rm d}=-D_{\rm CR}\vec{b}(\vec{b}. \nabla) e_{\rm CR}$ is the anisotropic diffusion flux (where $\vec b$ is the magnetic field unit vector), $\vec{F}_{\rm st}=f_{\rm b,st}\vec{u}_{\rm st}(e_{\rm CR}+P_{\rm CR})$ is the streaming advection term, and $-\vec{u}_{\rm{st}} . \nabla P_{\rm CR}$ is the loss term due to Alfv\'en wave damping through streaming of the CR energy (directly transferred into the thermal ion pool of energy).
The streaming velocity is $\vec{u}_{\rm st}=-{\rm sign}(\vec{u}_{\rm A}.\nabla e_{\rm CR})\vec{u}_{\rm A}$, with $\vec{u}_{\rm A}=\vec{B}/\sqrt{4\pi \rho}$ the Alfv\'en velocity, and where $f_{\rm b,st}$  is a streaming velocity boost term that depends on the strength of the turbulent and non-linear Landau wave damping in the hot ICM~\citep{Wiener2013}. This is typically about unity for galaxy clusters~\citep{Ruszkowski2017}. For the simulations presented here, $f_{\rm b,st}=1$.
We set the CR diffusion coefficient to $D_{\rm CR}=10^{29} \rm \ cm^2 s^{-1}$, which is the typical value obtained in simulated galaxies to match the observed \gammaray~ flux~\citep{Chan2019,Hopkins2021}, or obtained with detailed models of CR propagation in the Milky Way~\citep[e.g.][]{Trotta2011}. To ensure that  our results do not sensitively depend on this choice, we test other values of $D_{\rm CR}$ and $f_{\rm b,st}$ in Appendix \ref{sec:varCR}.

The evolution of the CR energy density can be more accurately described by taking the first two moments of the Vlasov equation, which augments the description of the CR evolution with a partial differential equation on the time evolution of the CR flux (as in \citealp{Jiang2018}, and in the more complete derivation of \citealp{Thomas2019}, who  also evolved the energy density of Alfvén waves with their damping processes). In the limit of the steady state for the CR flux (i.e. when the speed of light tends towards infinity), the two-moment method is equivalent to the standard approach considered here (and commonly employed in the literature; e.g.~\citealp{Ruszkowski2017, Girichidis2018, Holguin2019, Dashyan2020, Buck2020, Semenov2021}). While these two-moment methods are appealing and show different behaviours in some tailored test problems (see e.g. the CR over-density test case on top of a strong background CR density of~\citealp{Thomas2019}), they usually rely on a reduced value of the speed of light for the simulations to remain feasible, which needs to remain sufficiently high to guarantee convergence to the correct solution. Nonetheless, \cite{Chan2019} have shown in the context of Milky Way-like simulations that the two methods lead to essentially equivalent results.

Electron energy was treated in a separate equation (eq.~\ref{eq:electron}) from the total energy (eq.~\ref{eq:total}) in order to take local thermo-dynamical equilibrium (LTE) effects into account.
The term $\vec{F}_{\rm cond}=(-\kappa_{\rm e}\vec{b}(\vec{b}. \nabla) T_{\rm e})$ is the anisotropic conductive flux, where 
\begin{equation}
\kappa_{\rm e}=f_{\rm sat}\kappa_{\rm sp}=f_{\rm sat} n_{\rm e}k_{\rm B} D_{\rm cond}
\end{equation}
is the Spitzer conductivity for electrons with $n_{\rm e}=\rho/(\mu_{\rm e} m_{\rm p})$ (and $n_{\rm i}=\rho/(\mu_{\rm i} m_{\rm p})$) the electron (ion) number density, $\mu_{\rm e}$ and $\mu_{\rm i}$ the electron and ion mean molecular weights, $m_{\rm p}$ the proton mass, $k_{\rm B}$ the Boltzmann constant, and $D_{\rm cond}$ the thermal diffusivity, which is equal to 
\begin{align}
    D_{\rm cond}&=1.31\lambda_{\rm e}\left(\frac{k_{\rm B} T_{\rm e}}{m_{\rm e}}\right)^{\frac{1}{2}} \nonumber \\
    &\simeq8.3\times 10^{30} \left(\frac{T_{\rm e}}{10^8 \,\rm K}\right)^{\frac{5}{2}}
    \left(\frac{n_{\rm e}}{10^{-2} \,\rm cm^{-3}}\right)^{-1} \, \rm cm^2\,s^{-1}\, ,
\end{align}
with the mean free path of electrons
\begin{align}
    \lambda_{\rm e}&=\frac{3^{3/2} (k_{\rm B}T_{\rm e})^2} {4 \pi^{1/2} n_{\rm e} q_{\rm e}^4 \ln \Lambda}\nonumber \\
     &\simeq2.1 \left(\frac{T_{\rm e}}{10^8 \,\rm K}\right)^{2}
    \left(\frac{n_{\rm e}}{10^{-2} \,\rm cm^{-3}}\right)^{-1}   \, \rm kpc\, ,
\end{align}
where $m_{\rm e}$ is the mass of the electron, $q_{\rm e}$ the charge of the electron, and $\ln{\Lambda}\simeq40$ the Coulomb logarithm.
This conductive flux saturates when the characteristic scale length of the gradient of temperature $\ell_{T_{\rm e}}=T_{\rm e}/\nabla T_{\rm e}$ is comparable to or shorter than the mean free path of electrons~\citep{Cowie1977}.
We followed~\citet{Sarazin1986} and introduced an effective conductivity that approximates the solution in the unsaturated and saturated regime by
\begin{equation}
\label{eq:sat}
    f_{\rm sat}=\frac{1}{1+4.2 \lambda_{\rm e}/\ell_{T_{\rm e}}}\, .
\end{equation}

The electron energy equation has an additional term $Q_{\rm e \leftrightarrow i}$ that transfers heat between electrons and ions and is responsible for restoring LTE,
\begin{equation}
Q_{\rm e \leftrightarrow i}= \frac{ T_{\rm i}-T_{\rm e}} {\tau_{\rm eq, ei}} \frac{n_{\rm e}k_{\rm B}} {\gamma-1} \, ,
\end{equation}
with the equilibrium timescale
\begin{align}
\tau_{\rm eq, ei}&=\frac{3 m_{\rm e}m_{\rm p}} {8 (2 \pi)^{1/2} n_{\rm i} q_{\rm e}^4 \ln \Lambda} \left( \frac{k_{\rm B} T_{\rm e}} {m_{\rm e}} \right)^{\frac{3}{2}} \nonumber \\
&\simeq20 \left(\frac{T_{\rm e}}{10^8 \,\rm K}\right)^{\frac{3}{2}}
    \left(\frac{n_{\rm i}}{10^{-2} \,\rm cm^{-3}}\right)^{-1} \,\rm Myr\, .
\end{align}

The radiative loss term for the total energy density $\mathcal{L}_{\rm rad}=\mathcal{L}_{\rm th}+\mathcal{L}_{\rm CR}+\mathcal{H}_{\rm CR\rightarrow th}$ contains the gas radiative loss term $\mathcal{L}_{\rm th}$ (see Section~\ref{sec:cooling}),
and the CR radiative loss term due to Coulomb and hadronic collisions~\citep{Guo2008} $\mathcal{L}_{\rm CR}=-7.51\times 10^{-16}(n_{\rm e}/{\rm cm^{-3}})(e_{\rm CR}/{\rm erg\, cm^{-3}})\,\rm erg\,s^{-1}$ reprocessed into the thermal pool at a rate $\mathcal{H}_{\rm CR\rightarrow th}=2.63\times 10^{-16}(n_{\rm e}/{\rm cm^{-3}})(e_{\rm CR}/{\rm erg\, cm^{-3}})\,\rm erg\,s^{-1}$

The induction equation was solved with the constrained transport scheme~\citep{Teyssier2006} that guarantees $\nabla.\vec{B}=0$ at machine precision.
The MHD system of equations was first solved by zeroing all source terms on the right-hand side of the equations using the MUSCL-Hancock scheme~\citep{Fromang2006} with linear reconstruction of the conservative variable, a minmod total variation diminishing scheme, and the HLLD approximate Riemann solver~\citep{Miyoshi2005}. For the hydrodynamics-only simulations we used for comparison, the HLLC Riemann solver~\citep{Toro2009} was used instead. For the anisotropic thermal conduction and CR diffusion, their corresponding fluxes were solved with an implicit method~\citep{Dubois2016}, and similarly for the advection part of the streaming, where it was treated just as a diffusion-like term~\citep{Dubois2019}.
As in~\cite{Dashyan2020}, the diffusion solver included a minmod slope limiter on the transverse component of the face-oriented flux that preserved the positivity of the solution~\citep{Sharma2007}. 
The solution of the temperature coupling was obtained with an implicit update~\citep{Dubois2016}.

 \subsection{Refinement}
 \label{sec:refinement}
 
The simulations were performed in a box size of 8.7 Mpc with a root grid of $64^3$ cells (refinement level 6) and a corresponding coarse resolution of 136 kpc, which was adaptively refined by eight more levels up to a maximum resolution of $\Delta x=531$ pc (level 14). Refinement proceeded according to one of three refinement criteria: i) Cells were refined when their gas mass exceeded $[27089, 8713, 4098, 1621, 461, 152, 59, 12, 12] \times 1.47 \times 10^6 \msun$  for levels 6 to 14 . ii) All cells within a sphere of radius 4$\Delta x$ radius from the black hole (BH) were maximally refined at all times. iii) To avoid the common problem of mass-based refinement criteria, which derefine and disperse low-density jet bubbles, we ensured that the AGN jet and its cavities remained refined by using a dedicated passive scalar. This scalar was introduced to the simulation by setting $f_{\rm jet}=1$ for all cells accelerated at the base of the jet (See Section \ref{sec:setup_AGN} for details). It was then advected with the flow and mixed into the background medium as the jet evolved. All cells whose value of the passive scalar injected at the AGN jet base exceeded $f_{\rm jet} = \rho_{\rm scalar}/\rho_{\rm gas}>10^{-4}$ were refined when the local gradient of this passive  scalar exceeded 10\%. To avoid filling the entire box with jet scalar over time and to ensure that the highest value of $f_{\rm jet}$ always marked the jet nozzle and recent jet bubbles, the passive jet scalar exponentially decayed with a decay timescale of 10 Myr. The combination of these refinement criteria ensured that dense collapsing regions and regions that were recently affected by AGN feedback, including hot low-density bubbles that would derefine under a purely Lagrangian refinement scheme, remained well refined throughout the simulation.

\subsection{Initial conditions of gas and dark matter}
\label{sec:initial_conditions}

\begin{figure}
    \centering
    \includegraphics[width=\columnwidth]{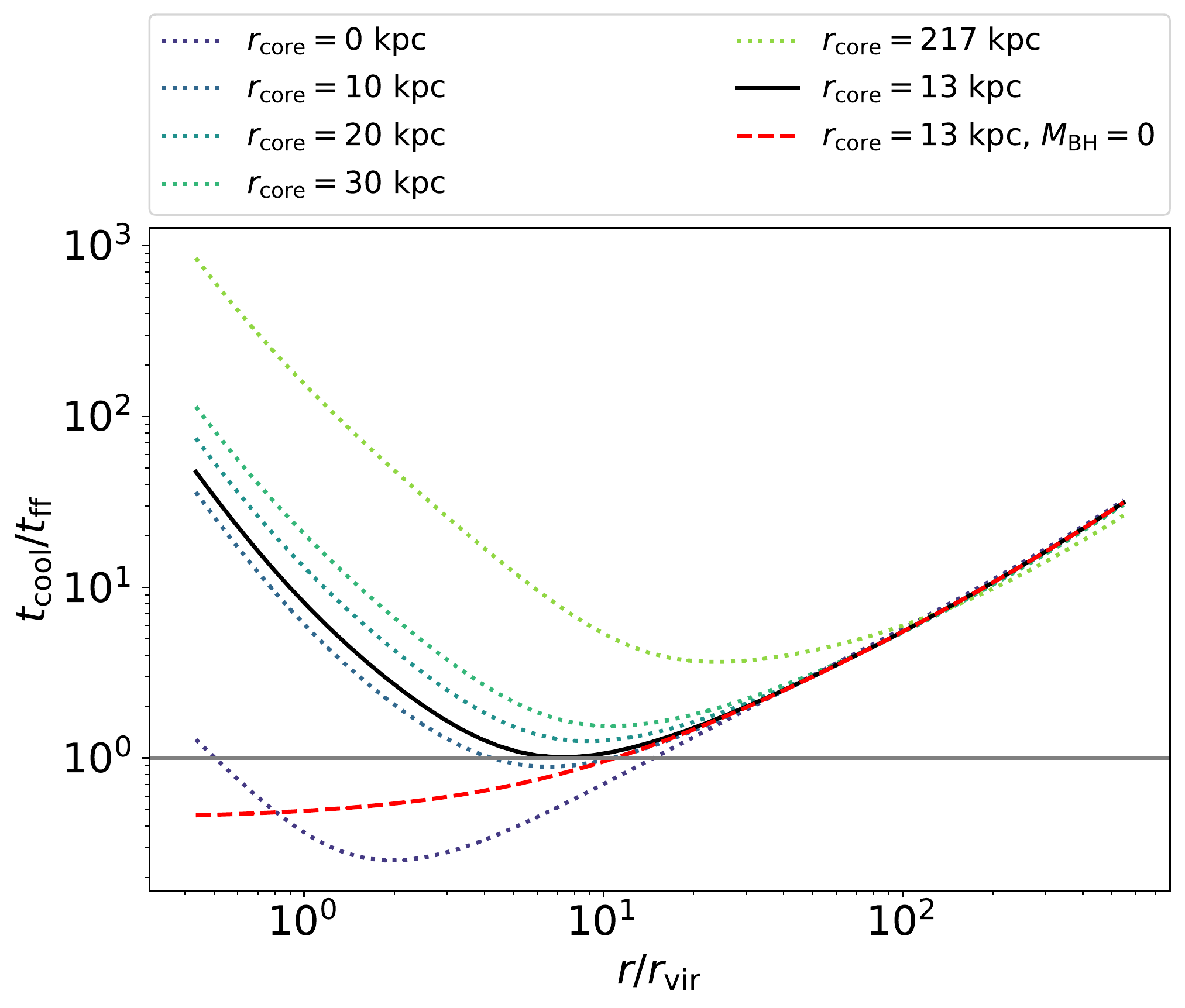}
    \caption{Ratio of the cooling time to the free-fall time for the initial conditions of the cluster for various values of the core radius $r_{\rm core}$ assuming an BH with a mass of $M_{\rm BH} = 1.6 \times 10^{10} \,\msun$. Without the mass of the BH, the initial cluster profile dips lower in the centre for the chosen core radius of 13 kpc (red line).}
    \label{fig:ini_tff_profile}
\end{figure}

The cluster was initialised as a cored NFW profile with a total mass of 
\begin{equation}
\rho_{\rm NFW}=\rho_{\rm s}\frac{r_{\rm s}^3}{(r+r_{\rm core})(r+r_{\rm s})^2} \, , 
\label{eq:NFW}
\end{equation}
where $r_{\rm s}=r_{200}/c$ is the scale radius, $r_{\rm core}=13 \rm \ kpc$ is the core radius, $\rho_{\rm s}=\rho_{\rm c}\delta_{200}$ is the density scaling of the profile, with the rescaling factor $\delta_{200}=200\Omega_{m}c^3/(3f(c))$, where $f(c)=\ln(1+c)-c/(1+c),$ and $\rho_{c}$ is the critical density of the Universe. The total mass of the halo was split into a fixed dark matter (DM) profile, which has the form $\rho_{\rm DM} = (1-f_{\rm gas}) \rho_{\rm NFW}$, and a gas profile of the form $\rho_{\rm gas} = f_{\rm gas} \rho_{\rm NFW}$ , where $f_{\rm gas}$ is the initial gas fraction. The DM profile was fixed and did not evolve throughout the simulation, whereas the gas profile did evolve under gravity, cooling, and the influence of the AGN jet. 

For the cluster presented here, which had a total mass of $8 \times 10^{14} \ \msun$, the parameters in the NFW profile were as follows: The concentration parameter was chosen to be $c_{200}=4.41$ based on the halo mass according to equation 9 of \citet{Maccio2007}  for relaxed halos. The combined choice of halo mass and concentration parameter leads to a scale radius $r_{\rm s}=434 $ kpc, a virial radius of $r_{200} = 1917$ kpc, and a virial velocity of $u_{200} = 1344 \,\rm km\,s^{-1}$. The gas fraction was chosen according to equation 5 of \citet{Andreon2017}, based on the cluster mass. For the cluster presented here, it is equal to $f_{\rm gas} = 0.103$. CR pressure was initialised based on the gas pressure profile using a low and constant value of $\eta = 10^{-4}$.

To ensure both local and global thermal stability in the initial conditions, the core radius was chosen such that the ratio of the cooling time $t_{\rm cool}$ was greater than the free-fall time $t_{\rm ff}$ in the cluster centre, which for the current parameters leads to a core radius of $r_{\rm core}=13 \, \rm kpc$. As can be seen by comparing the dashed red line ($r_{\rm core}=0\,\rm kpc$ and $M_{\rm BH}=0\, \rm M_\odot$) to the dotted green lines ($r_{\rm core}\ne 0\,\rm kpc$ and $M_{\rm BH}=1.6\times 10^{10}\, \rm M_\odot$) in Fig. \ref{fig:ini_tff_profile}, the gravitational potential of the BH is crucial in producing an upturn in $t_{\rm ff}/t_{\rm cool}$ in the cluster centre. Without it, extremely large core radii would be needed to meet the thermal stability criterion.

The cluster profile was truncated at the virial radius. Outside of $r_{200}$, the density scales as
\begin{equation}
    \rho(r>r_{200}) = \rho(r_{200}) \left(\frac{r_{200}}{r}\right)^5
,\end{equation}
where $\rho(r_{200})$ is the density at $r_{200}$ according to Eq. \ref{eq:NFW}.

Avoiding a sharp density cutoff at the outer cluster edge has several advantages. It increases stability in the cluster outskirts, avoiding large-scale bulk flows due to poorly resolved hydrostatic equilibrium in areas of rapid density transition, and it considerably speeds up calculations by reducing conductive heat flows because sharp temperature transitions are smoothed. As this study is focused on the cluster centre, the choice of parameters for the cluster edge has a negligible impact on the  results of this study, but it makes an important difference to the overall computational resources required, with parameters here chosen to minimise computational requirements.

Ions and electrons were initial in LTE. The initial ion pressure was set to
\begin{equation}
    P_{\rm i}=P_{\rm e} \frac{\mu_{\rm i}}{1+\mu_{\rm e}}\, , 
\end{equation}
where the mean molecular weights were $\mu_{\rm i}=1.22$ and $\mu_{\rm e}=1.14$ for a combined gas mean molecular weight of $\mu_{\rm gas}= 0.59$. Hydrostatic equilibrium in the initial conditions was calculated so that the total (thermal, CR, and magnetic) pressure gradient balanced the gravitational pull.

\subsection{Initial conditions of the magnetic field}

All magnetised simulations were initialised with a three-dimensional randomly tangled magnetic field on a $512^3$ grid. In order to guarantee $\nabla.\vec B=0$, we set up a random magnetic potential vector $\vec A$ at each cell corner, and reconstructed the $\vec B$ field at the centre of cell faces by taking the curl of $\vec A$. This initial magnetic field was then resampled onto the initial adaptive grid using a procedure that enforced $\nabla.\vec B=0$ at machine precision even at refined interfaces. For the box size of 8.7 Mpc used here, this means that the magnetic field was initialised with a characteristic coherence length of 17 kpc. As the simulation progressed, this coherence length evolved in accordance with local gas flows. 

The magnetic field was initialised with a central magnetic field strength of $B_0=20 \rm \ \mu G$ and was then scaled with the initial density profile as $B(r) = B_0 (\rho(r)/\rho_0)^{2/3}$. Due to the tangled nature of the initial field, numerical magnetic reconnection took place, which somewhat decreased the average magnetic field strengths throughout the simulation.

\subsection{Cooling, metallicity, and star formation}
\label{sec:cooling}

Radiative cooling of the gas was calculated according to the tabulated values of \citet{Sutherland1993} for temperatures above $T>10^4 \,\rm K$, with values extended below $10^4$ K using the fitting functions from \citet{Rosen1995}. For metal cooling, the local metallicity value of each cell was taken into account, with solar abundance ratios of elements assumed throughout. 

\begin{figure}
    \centering
    \includegraphics[width=\columnwidth]{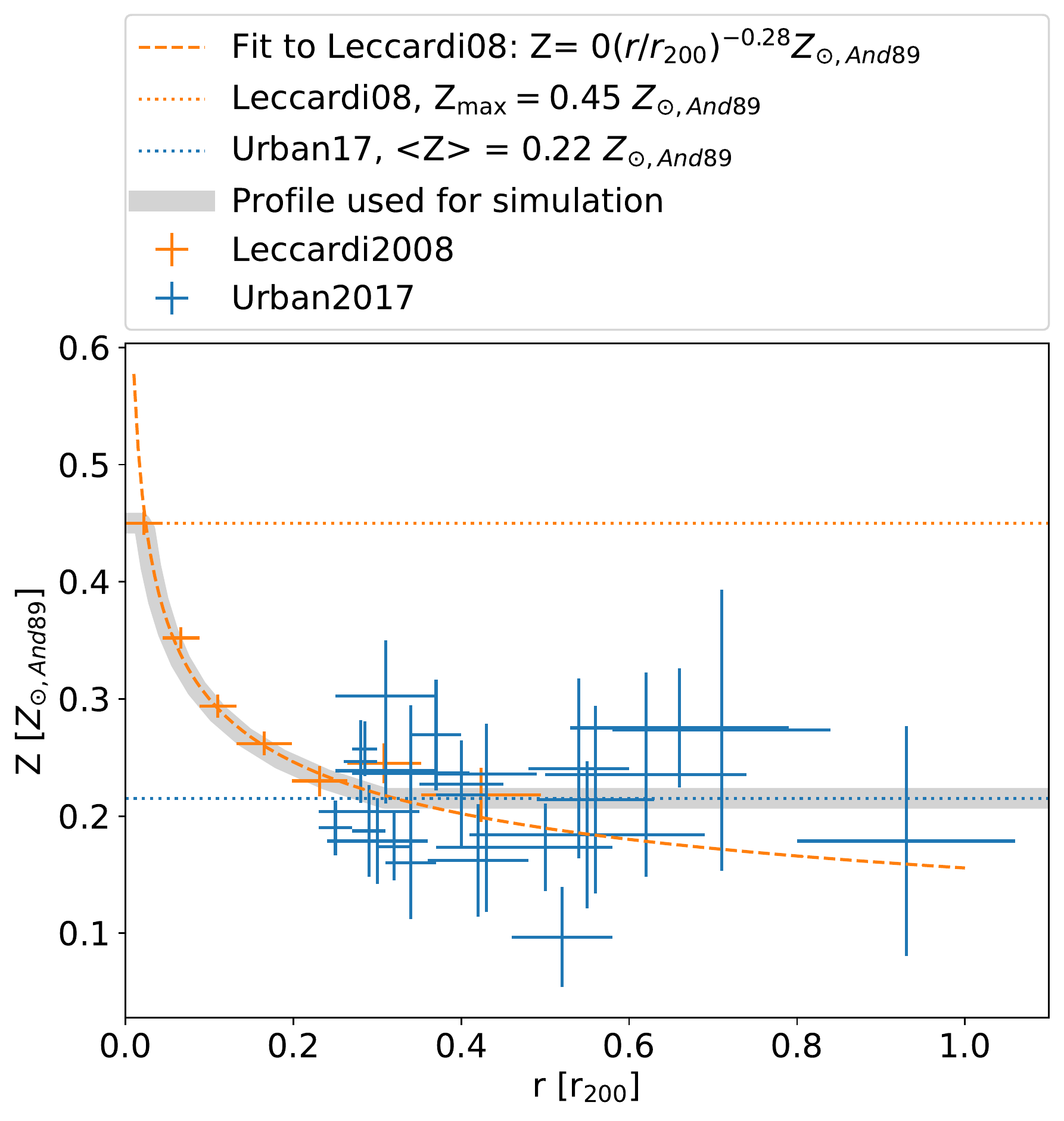}
    \caption{Metallicity profile used for the cluster, based on observational data by \citet{Leccardi2008} and \citet{Urban2017} in units of solar metallicity according to \citet{Anders1989}}
    \label{fig:z_profile}
\end{figure}

Metals were treated as a single species and were advected as a passive scalar. The cluster was initialised with a metallicity profile as shown in Fig. \ref{fig:z_profile}, which was computed according to 
\begin{equation}
    Z=\min\left(0.45,\max\left(0.22,0.15 \left(\frac{r}{r_{200}}\right)^{-0.28}\right)\right)
,\end{equation}
with the upper limit and fit taken from \citet{Leccardi2008}, an the lower limit being the average value of data reported in \citet{Urban2017}. The metallicity profile we used is therefore flat in the cluster centre, falls off to about $0.3 r_{\rm 200}$ , and is flat outside this region. Gas was further metal-enriched throughout the simulation due to supernova explosions.

Stars were formed according to combined density and temperature criteria, with star formation permitted in cells with a hydrogen number density of $n_{\rm H} > 0.1 \, \rm  H\, cm^{-3}$ and a temperature $T < 10^4\,\rm K$. This led to a stellar mass resolution of $m_{\rm *,min}=n_{\rm H} m_{\rm p} \Delta x^3 / X_{\rm H} = 3.89\times 10^5\, \msun$ , where $m_{\rm p}$ is the proton mass, and $X_{\rm H}= 0.74$ is the fractional abundance of hydrogen. Star formation produces stellar particles, randomly drawn with a Poisson process~\citep{Rasera2006}, following a Schmidt law $\dot \rho_* = \epsilon_* \rho /t_{\rm ff}$ , where $\rho$ is the gas density, $t_{\rm ff}$ is the free-fall time of the gas cell, and $\epsilon_*=0.1$ is the constant efficiency of star formation.

Stellar feedback was included in the form of type II supernovae only, using the energy-momentum model of \citet{Kimm2015}. Each stellar particle releases a total energy of $e_{\rm *,SN} =  m_* \eta_{\rm SN} 10^{50} \,\rm erg\,\msun^{-1}$ in one feedback event after 10 Myr, where $\eta_{\rm SN} = 0.2$ corresponds to the mass fraction of the initial mass function for stars ending their life as type II supernovae, and $m_*$ is the stellar particle mass. These explosions also enriched the gas locally with metals with a constant yield of 0.1.

\subsection{Active galactic nucleus}
\label{sec:setup_AGN}

To model the self-regulating AGN in the centre of the cluster, an BH was placed at the centre of the cluster as part of the initial conditions. This BH had a mass of $M_{\rm BH}= 1.65 \times 10^{10} \msun$, following the $M_{\rm BH} - M_{500}$ relation in \citet{Phipps2019}. 

\subsubsection{Dynamics}
\label{sec:gradientdescent}

The BH was modelled as a \textsc{ramses} sink particle, and it was free to move within the gravitational potential of the halo. To compensate for unresolved dynamical friction, we employed two related sub-grid algorithms: we calculated the dynamical friction due to gas according to \citet{Ostriker1999}, using the high-resolution correction from \citet{Beckmann2018} to prevent the dynamical friction sub-grid algorithms from ejecting the BH when local overdensities are resolved, and the dynamical friction due to the stars~\citep{Pfister2019}.

As the initial conditions did not include the stellar component of the central galaxy and the DM halo has a strong core, the gravitational potential in the cluster centre was very shallow at the beginning of the simulation. This caused the BH to wander occasionally from the centre of the galaxy cluster. To diminish this effect, we added a gradient descent correction of the form 
\begin{equation}
    \mathbf{x}_{n+1} = \mathbf{x}_{n} - \tilde{\gamma}_d \mathbf{f}(\mathbf{x}_n)
\end{equation} to the BH trajectory according to Pellissier (2022, in prep.). Here, $\mathbf{x}_n$ is the position of the BH at step n, and $\mathbf{f}(\mathbf{x}_n) = \nabla \phi$ is the gradient on the gravitational potential $\phi$. $\tilde{\gamma}_d = f_{\rm grad} \sqrt{\gamma_d} \Delta t$, where $f_{\rm grad}$ is a dimensionless pre-factor to scale the magnitude of the acceleration, $\Delta t$ is the time step of the simulation, and \begin{equation}
\gamma_d = \frac{|(\mathbf{x}_{n+1} - \mathbf{x}_{n})^T[\mathbf{f}(\mathbf{x}_{n+1}) - \mathbf{f}(\mathbf{x}_{n})]|}{||\mathbf{f}(\mathbf{x}_{n+1}) - \mathbf{f}(\mathbf{x}_{n}) ||^2}
\end{equation}
following \citet{Barzilai1988}. This adds a small force along the steepest local gradient descent, helping the BH to remain attached to the centre of the cluster. The contribution of the gradient descent acceleration can be controlled with the free parameter $f_{\rm grad}$.

\subsubsection{Accretion and spin} 
\label{sec:accretion_spin}

The BH accretes according to the Bondi-Hoyle-Lyttleton (BHL) accretion rate, capped at a maximum of 1\% the Eddington accretion rate,
\begin{equation}
    \dot{M}_{\rm BH} = \min(\dot{M}_{\rm BHL},0.01 \dot{M}_{\rm Edd})
,\end{equation}
where
\begin{equation}
    \dot{M}_{\rm BHL} = \frac{4 \pi (G M_{\rm BH})^2 \bar{\rho}}{(\bar{c}_s^2+\bar{v}_{\rm BH,rel}^2)^{3/2}}
    \label{eq:dotMBHL}
\end{equation}
is the Bondi-Hoyle-Lyttleton accretion rate, and 
\begin{equation}
    \dot{M}_{\rm Edd} = \frac{4 \pi G M_{\rm BH} m_{\rm p}}{\epsilon_{\rm mad} c \sigma_{\rm T}}
\end{equation}
is the Eddington accretion rate. $G$ is the gravitational constant, $\bar{\rho}$, $\bar{c}_s$ , and $\bar{v}_{\rm BH,rel}$ are the weighted local average density, sound speed, and relative velocity between BH and gas, respectively~\citep[see][for details]{Dubois2012}. $m_{\rm p}$ is the proton mass, $c$ is the speed of light,  $\sigma_{\rm T}$ is the Thompson cross section, and $\epsilon_{\rm MAD}(a_{\rm BH})$ is the spin-dependent feedback efficiency from \citet{Dubois2021} based on MAD simulations by \citet{McKinney2012}. 

All quantities were measured within a radius $4 \Delta x$ of the BH location. Throughout the evolution of the BH, its spin vector $\vec{a}_{\rm BH}$ was followed self-consistently through accretion~\citep[see][for technical details]{Dubois2014,Dubois2021}. The BH was initialised as non-spinning, and wa spun up or down according to the angular momentum of the accreted gas throughout the simulation. The spin magnitude and orientation were updated assuming a magnetically arrested disc~\citep[MAD;][]{McKinney2012} at all times.

\subsubsection{Feedback}
\label{sec:AGN_feedback}

\begin{table*}
    \centering
    \caption{Simulation properties for simulations with an AGN. All simulations have a total cluster mass of $8 \times 10^{14}\,\msun$ and an BH with a mass of originally $1.6 \times 10^{10}\, \msun$. All MHD simulations treat electrons and ions separately, while the non-MHD simulation (HYDRO) has only one temperature (and no CR). $f_{\rm grad}$ stands for the strength of the gradient descent for the BH (see Section~\ref{sec:gradientdescent}). The asterisk in CRc\_dsh\_f1\_2$^*$ indicates that the sub-grid drag force from gas and stars is turned off.}
    \begin{tabular}{c c c c c c c c}
         \bf Simulation   & \bf Initial & \bf Jet energy &  \bf $f_{\rm grad}$ & \bf  Thermal & \bf CR & \bf CR streaming & \bf CR streaming    \\
         \bf name & \bf B field & \bf fraction &  & \bf  conduction & \bf diffusion & \bf advection & \bf heating   \\
         \hline
        HYDRO & no & 100\% kinetic  & 0.2 & no & no & no & no  \\
         \hline
         MHDc & yes & 100\% kinetic & 0.2 & yes & no & no & no \\
        \hline
        CRc\_dsh & yes & 10\% kinetic, 90\% CR & 0.2 & yes & yes & yes & yes \\
        CRc\_dsh\_weakcr & yes & 90\% kinetic, 10\% CR & 0.2 & yes & yes & yes & yes\\
        \hline
        CRc & yes & 10\% kinetic, 90\% CR & 0.2 & yes & no & no & no \\
        CRc\_d & yes & 10\% kinetic, 90\% CR & 0.2 & yes & yes & no & no \\
        CRc\_s & yes & 10\% kinetic, 90\% CR & 0.2 & yes & no & yes & no \\
        CRc\_h & yes & 10\% kinetic, 90\% CR & 0.2 & yes & no & no & yes \\
        \hline
        CRc\_ds & yes & 10\% kinetic, 90\% CR & 0.2 & yes & yes & yes & no \\
        CRc\_dh & yes & 10\% kinetic, 90\% CR & 0.2 & yes & yes & no & yes \\
        CRc\_sh & yes & 10\% kinetic, 90\% CR & 0.2 & yes & no & yes & yes \\
        \hline
         CRc\_dsh\_f0.5 & yes & 10\% kinetic, 90\% CR & 0.5 & yes & yes & yes & yes \\
         CRc\_dsh\_f1 & yes & 10\% kinetic, 90\% CR & 1 & yes & yes & yes & yes \\
        CRc\_dsh\_f1\_2$^*$ & yes & 10\% kinetic, 90\% CR & 1 & yes & yes & yes & yes \\
        \hline
        CRc\_dsh\_1 & yes & 10\% kinetic, 90\% CR & 0.2 & yes & yes & yes & yes \\
        CRc\_dsh\_2 & yes & 10\% kinetic, 90\% CR & 0.2 & yes & yes & yes & yes \\
        \hline
    \end{tabular}
    \label{tab:AGN_simulations}
\end{table*}

We modelled AGN feedback with jets following the method presented in \citet{Dubois2012}. At each time step of the simulation, the AGN has a luminosity of
\begin{equation}
    L_{\rm AGN} = \epsilon_{\rm MAD}(a_{\rm BH}) \dot{M}_{\rm BHL} c^2
    \label{eq:LAGN}
.\end{equation}
This luminosity was used to accelerate gas within a jet cylinder of width of $2\Delta x_{\rm min}$ and a height of $4 \Delta x_{\rm min}$, whose major axis was (anti)aligned with the BH spin vector. Jets were injected with a velocity of $10^5 \rm \ km s^{-1}$. As this spin vector naturally evolved throughout the simulation under the influence of the precipitation of cold gas onto the cluster centre and its chaotic accretion onto the BH \citep{Gaspari2013}, there was no need to add an ad hoc precession to produce several generations of inflated AGN bubbles \citep{Beckmann2019b}. At an average density ratio of $\rho_{\rm jet} / \rho_{\rm ICM} = 0.03$, our simulation modelled light jets.

The energy deposited in the jet was split into two channels: kinetic energy $L_{\rm kin}= (1-f_{\rm CR}) L_{\rm AGN}$ , and CR energy $L_{\rm CR} = f_{\rm CR} L_{\rm AGN} $ according to a fixed CR energy fraction $f_{\rm CR}$. 

\subsection{Simulation parameters}

We present a suite of simulations to explore the impact of individual physical mechanism such as magnetic fields and CRs and their associated choice of parameters in a comparative manner. A summary of the parameters we used for all simulations can be found in Table \ref{tab:AGN_simulations}.

\section{Impact of magnetic fields and cosmic rays on the intracluster medium}  
\label{sec:suite}


\begin{figure*}
    \centering
    \includegraphics[width=\textwidth]{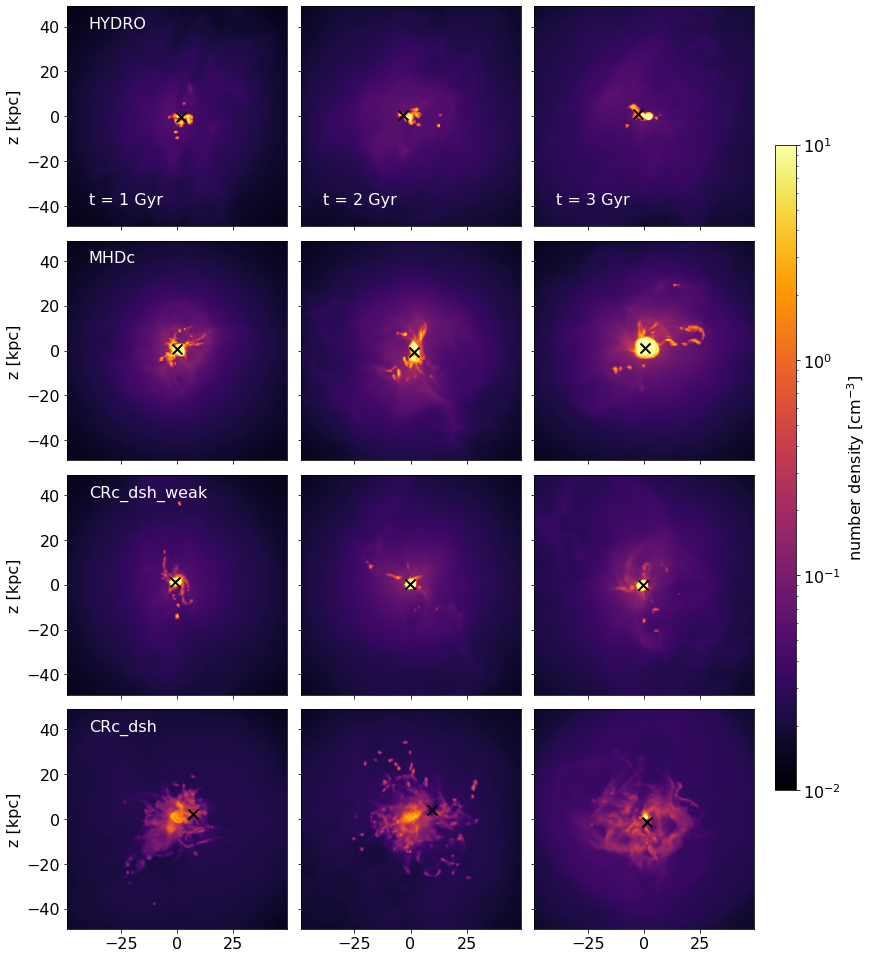}
    \caption{Projected gas density showing the multi-phase structure of the gas in the cluster centre for different physics (rows) and at three different times 1, 2, and 3 Gyr from left to right. The location of the BH is marked in black. The morphologies and the extent of the condensed gas depend strongly on the presence of magnetic fields and CRs.}
    \label{fig:suite_images}
\end{figure*}

Adding magnetic fields and CRs has a significant and persistent impact on the amount and distribution of warm and cold gas in the cluster centre. This is shown in Fig.~\ref{fig:suite_images}. 

\begin{figure}
    \centering
    \includegraphics[width=\columnwidth]{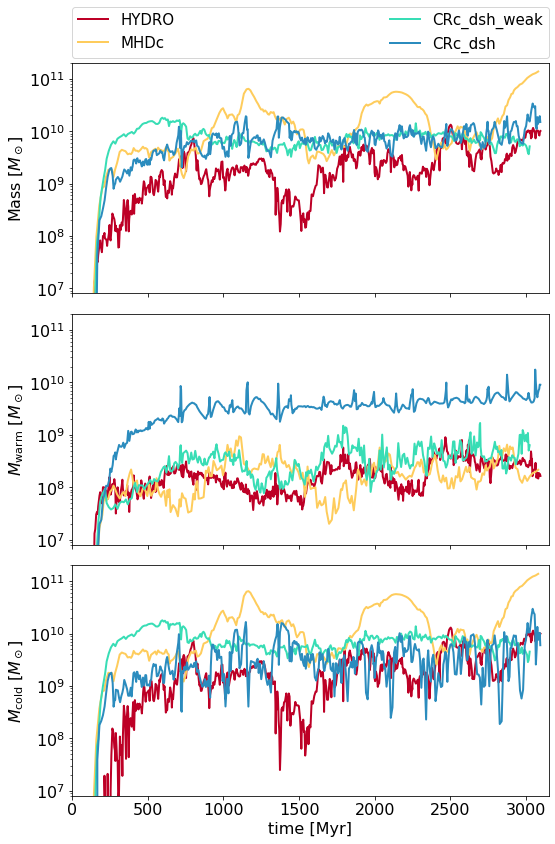}
    \caption{Time evolution of the total gas mass of all gas that has $T<10^6\, \rm K$ ($M_{\rm warm}+M_{\rm cold}$, top panel), the warm gas mass ($M_{\rm warm}$, $10^4 \, {\rm K}<T<10^6 \, \rm K$, middle panel), and the cold gas mass ($M_{\rm cold}$, $T<10^4 \,\rm K$, bottom panel) in the galaxy cluster under the influence of different physical mechanisms as indicated in the top inset. With weak CRs, a cooling catastrophe is avoided, and with strong CRs, the bulk of the condensed gas is maintained in the warm phase for long periods of time.}
    \label{fig:suite_gas_quantities}
\end{figure}

\begin{figure}
    \centering
    \includegraphics[width=\columnwidth]{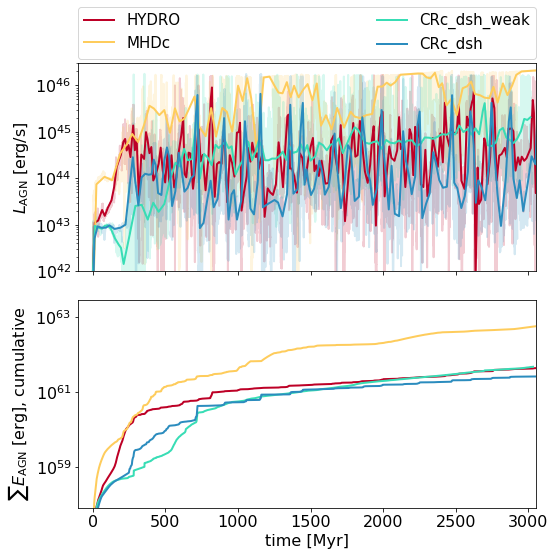}
    \caption{Time evolution of the AGN luminosity (top panel) and of the cumulative energy injected by the AGN (bottom panel) for the different simulations with colours as indicated in the top inset. Even a small fraction of CRs in the jet allows the AGN to maintain self-regulation of the cluster over long periods of time.}
    \label{fig:suite_AGN_timeseries}
\end{figure}

In the absence of magnetic fields and CRs, (run HYDRO), a dense, centralised clump forms early on, with little dense gas found beyond 5 kpc from the cluster centre at any point during the simulation. The timeseries in Fig. \ref{fig:suite_gas_quantities} shows that the AGN is able to regulate cooling in the cluster for the 3 Gyr of evolution probed here, with the central gas feature containing $10^9$ to $10^{10} \, \msun$ at all times. The AGN luminosity varies in the range $10^{43} - 10^{46} \, \rm erg\,s^{-1}$ on duty cycles of about 100 Myr (Fig. \ref{fig:suite_AGN_timeseries}).

With the addition of magnetic fields and anisotropic thermal conduction, MHDc, the central dense core remains, but locally thermally unstable regions extend farther from the cluster centre, as shown by the filamentary dense gas structure that surrounds the central core in Fig. \ref{fig:suite_images}. This increased cooling flow leads to large cold gas reservoirs that can exceed $10^{11} \msun$ (see Fig. \ref{fig:suite_gas_quantities}), although the AGN injects more than an order of magnitude more cumulative energy throughout the 3 Gyr of evolution than in the HYDRO case (see Fig. \ref{fig:suite_AGN_timeseries}), at $6.1\times 10^{62} \rm \ erg$ for MHDc in comparison to $4.4\times 10^{61} \rm \ erg$ for HYDRO. The largest cold gas reservoir, which builds up from 2.8 Gyr, continues to grow by the end of the simulation, which is a clear sign that the AGN has become overwhelmed and fails to self-regulate gas cooling in the cluster. As was shown in previous simulations, this run-away cooling is rarely brought back under control by the AGN \citep{Li2015}. A comparison of the MHDc and HYDRO shows the effect discussed in \citet{Ji2018} that magnetic fields enhance thermal instability in clusters (see Section \ref{sec:theory} for a discussion).

The evolution of MHDc also shows that despite the presence of a powerful AGN jet that continuously stirs the cluster centre, anisotropic thermal conduction is unable to prevent cooling flows in massive galaxy clusters. This is due to a strong heat-buoyancy instability~\citep{Quataert2008,Parrish2009} that preferentially aligns magnetic field lines in a tangential configuration and shuts of the radial heat flows required to offset cooling in the cluster centre. The impact of thermal conduction on cluster cooling flows is discussed in detail in the companion paper (\citet{Beckmann2022b}). 

When CRs are added to an AGN jet, the choice has to be made how much energy is placed in CRs rather than kinetic energy. We tested two values of $f_{\rm CR}$: a fiducial strong CR simulation with $f_{\rm CR}=0.9$, called CRc\_dsh, and a complimentary weak CR simulation with $f_{\rm CR}=0.1$, called CRc\_dsh\_weak. 

By injecting the majority of the jet energy as CRs, the AGN needs one-third less energy to regulate the cluster cooling flow over 3 Gyr (see Fig. \ref{fig:suite_AGN_timeseries}). As a result, after also taking the CR injections from SNe into account, which contribute 8.4\% and 1.7\% of the overall CR energy for CRc\_dsh\_weak and CRc\_dsh, respectively, the cluster under the influence of a CR-dominated jet receives a total CR energy that is roughly five times that of CRc\_dsh\_weak over 3 Gyr (at $2.6\times 10^{61} \rm \ erg$ for CRc\_dsh, and $5.2\times 10^{60} \rm \ erg$ for CRc\_dsh\_weak).

Simulation CRc\_dsh\_weak in Fig. \ref{fig:suite_images} shows that a small number of CRs does not significantly change the morphology of the dense gas. However, with CRs, the total cold gas mass remains below $10^{10} \, \msun$  (Fig. \ref{fig:suite_gas_quantities}) and the AGN continues to self-regulate the cluster for the whole 3 Gyr. While it is still possible that the AGN becomes overwhelmed at a later stage, this is unlikely as the very consistent amount of cold gas in the simulation over more than 2 Gyr and the low variability in the AGN luminosity over the same time span are strong signs for a stable configuration. Overall, CRc\_dsh\_weak closely resembles HYDRO in its evolution. The total AGN energy injected in both simulations differs by only 5\%. This suggests that for a given cluster, a given amount of energy is required for self-regulation, which is robust to small changes in physics (this is discussed further in Appendix \ref{sec:stochastic}).

The images in Fig. \ref{fig:suite_images} show that the morphology of the condensed gas is much more volume filling and diffuse in the presence of a large CR fraction. The bottom panel of Fig. \ref{fig:suite_gas_quantities} shows that the bulk of this extended nebula is in the warm phase, that is, it has a temperature in the range $10^4 - 10^6\,\rm K$. In comparison to the other simulation, which contains $10^8 - 10^9\, \msun$ of gas in this phase, the strong CR simulation CRc\_dsh consistently produces more than $10^9\, \msun$ of warm gas (middle panel), which is compensated for by a decrease in the cold gas mass (bottom panel). The total gas mass, however, is very similar for CRc\_dsh and CRc\_dsh\_weak on average, showing only a variation of 30 \% between the two simulations. Therefore, a comparable amount of gas condenses from the hot phase in both simulations, but a stronger fraction of CRs keeps this gas warm for long periods of time, rather than allowing it to cool to $10^4$ K or lower. 

\begin{figure*}[t]
    \centering
    \begin{tabular}{c|c}
        \includegraphics[width=0.5\textwidth]{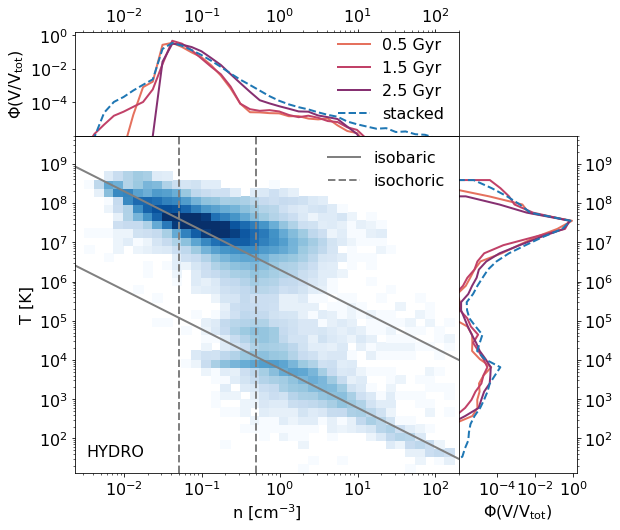} &
        \includegraphics[width=0.5\textwidth]{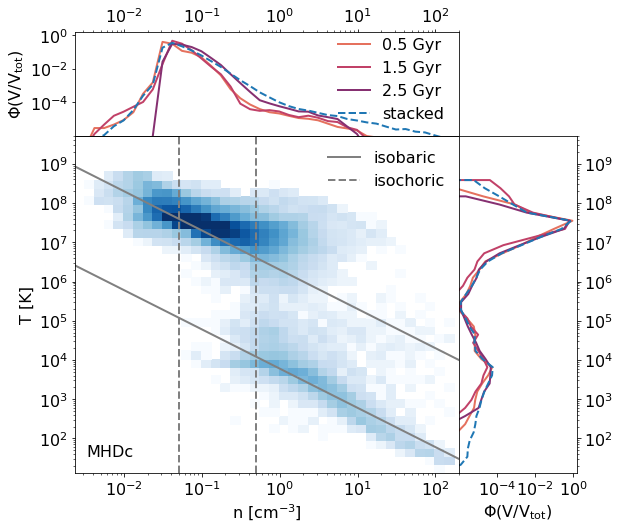}\\
        \includegraphics[width=0.5\textwidth]{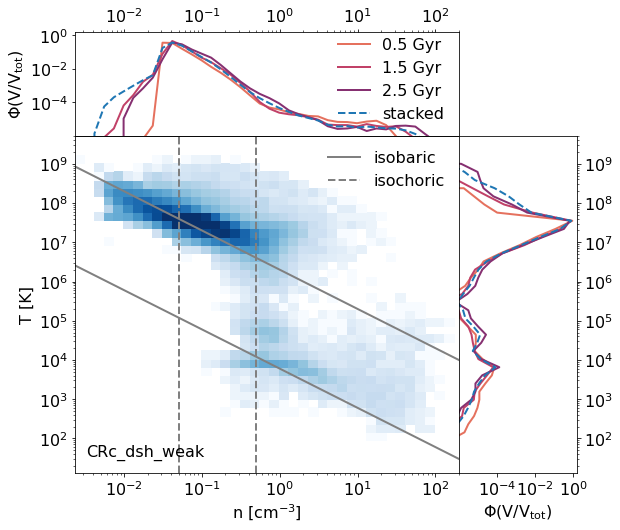} &   
        \includegraphics[width=0.5\textwidth]{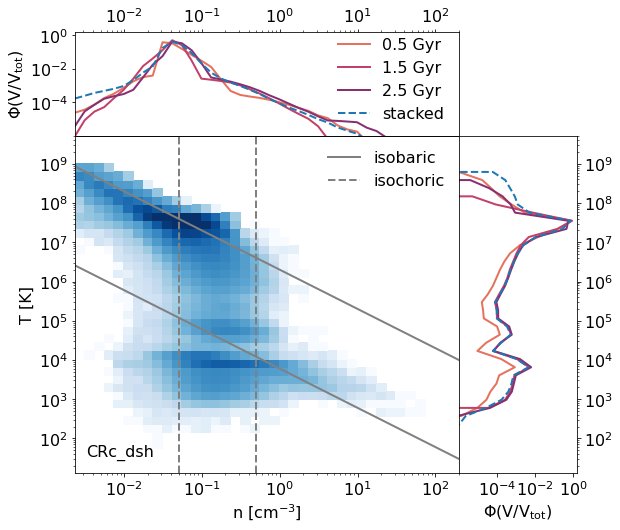}\\
    \end{tabular}
    \caption{2D phase plots showing the volume-weighted distribution of gas number density $n$ vs temperature $T$ within the central 60 kpc of the cluster for a stacked sample measured every 0.25 Myr between $t=0$ Gyr and $t=3$ Gyr. Solid (dashed) grey lines show example lines along which gas evolves when it cools in an isobaric (isochoric) fashion. Histograms along the x-axis (y-axis) show 1D histograms for the number density (temperature), weighted by $V/V_{\rm tot}$ , where $V_{\rm tot}$ is the total volume probed here. Solid lines show the distribution at individual points in time, and the dashed line shows the distribution for the stacked sample. With a CR-dominated jet, gas cooling in the cluster centre changes from isobaric (solid lines) to isochoric (dashed lines)}
    \label{fig:suite_phaseplot}
\end{figure*}

The impact of CR appears clearly in the phase plots shown in Fig. \ref{fig:suite_phaseplot}. In the HYDRO case, gas cools in a narrow isobaric fashion, except for a brief isochoric phase around $0.5 \rm \, cm^{-3}$. When it cools isobarically, gas evolves along a line with slope $T\propto n^{-1}$ in the plots shown here (solid line), while isochoric cooling causes it to evolve along a vertical line (dashed line). With the addition of magnetic fields, MHDc, the phases in which gas can be found do not significantly change in comparison to the non-magnetised case. With the addition of some CR, CRc\_dsh\_weak, the hot gas distribution narrows again, but the cold gas phase is broader. In general, there is no significant difference between properties stacked across different points in time (blue) versus individual snapshots (reds).

Only when a much larger fraction of the AGN luminosity is injected into CRs ($f_{\rm CR}=0.9$) does the distribution of the gas phases change significantly. As discussed in \citet{Kempski2019}, with the addition of a sufficient number of CRs, the cooling in the cluster becomes predominantly isochoric, with little isobaric evolution. This leads to the broad distribution of gas shown in CRc\_dsh. This isochoric cooling in the presence of CRs was also reported by \citet{Butsky2020} using simulations of stratified cooling boxes. This work also showed that the presence of CR turns cooling isochoric, but efficient CR transport again flattens the slope of $T$ versus $n$.

One consequence of this evolution is that with strong CR feedback, gas can be found in a warm and diffuse phase that is absent from any of the other simulation, that is, where $T<10^7\, \rm K$  and $\rm n<0.3 \, \rm cm^{-3}$ (grey box). As the cooling rate $\mathcal{L}_{\rm th}$ depends on the gas density, this warm and diffuse gas will cool much more slowly than gas at the same temperature but higher density. This leads to the buildup of warm gas in the cluster centre as discussed previously, and, hence, prevents the runaway cooling of the hot phase at $T\simeq 10^7-10^8 \,\rm K$ into the cold star-forming phase $T\ll 10^4 \,\rm K$. 

We caution that the isochoric cooling discussed here here might be partially or entirely due to the limited resolution of our simulations. As shown in \citet{Fielding2020}, low resolution produces artificial pressure dips during rapid cooling, which vanish as the structure of cooling gas becomes better resolved. It is possible that the transition to isochoric cooling reported here and in \citet{Butsky2020} is influenced by this lack of resolution, and that cooling would remain isobaric if individual cloudlets with radii of $< 1 \rm \ pc$ \citep{McCourt2018}, and their mixing layers, were fully resolved. This will remain technically impossible in cluster-scale simulations for the foreseeable future. We test the impact of resolution on cooling flows and phase transitions in our simulations in Appendix \ref{sec:resolution} and find little change within the accessible parameter space.

We conclude that if even a fraction as small as 10\% of the total jet energy is converted into CRs, the impact of magnetic fields on thermal instability is offset and AGN self-regulation of the cluster becomes more robust. Converting a larger fraction of AGN jet energy into CRs means that less overall jet energy is required to self-regulate the cluster cooling flow, but it also significantly changes the bulk cooling properties of the gas to isochoric. This leads to a broad distribution of warm diffuse gas in the cluster centre.  

\begin{figure}
    \centering
    \includegraphics[width=\columnwidth]{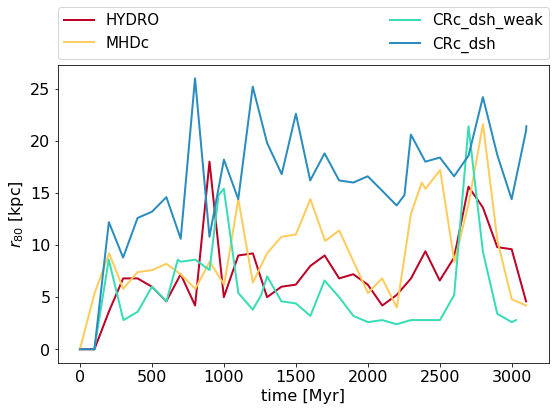}
    \caption{Time evolution of the radial extent of the warm gas from the cluster centre, measured as the radius $r_{80}$ that contains 80\% of the total gas mass with temperatures $T < 2 \times 10^5 $ K for the different simulations with different colours, as indicated in the top inset. With a CR-dominated jet, the central nebula is significantly more extended.}
    \label{fig:suite_r80}
\end{figure}

The images in Fig.~\ref{fig:suite_images} show that not only the morphology of condensed gas, but also its spatial distribution changes with increasing the CR fraction. As shown in Fig.~\ref{fig:suite_r80}, the radius that contains 80\% of the warm and cold gas ($T<2\times 10^5\,\rm K$), $r_{80}$, varies with time but is broadly similar for simulations HYDRO, MHDc, and CRc\_dsh\_weak. However, for CRc\_dsh, the bulk of the gas is found significantly farther from the cluster centre, out to radii as large as 20 kpc or more.

\section{Cosmic-ray transport mechanisms}
\label{sec:transport}

In this section, we investigate the role played by the three terms on the right-hand side of Eq. \ref{eq:CR}, that is, CR diffusion (``d'' for the letter standing in the simulation name after ``CRc\_''), CR redistribution through streaming advection (``s''), and transfer of energy from CR to thermal energy through streaming heating (``h''). For this investigation, we compared simulations with only a single term (e.g. CRc\_d with the diffusion term only) to simulations with two terms (e.g. CRc\_dh for diffusion and streaming heating) to the full-physics simulation CRc\_dsh and a simulation with CR, but without CR transport and related processes, CRc. We note that these reduced simulations have a limited physical validity as all transport mechanisms are physically tightly linked. They were performed here to gain insight into which term in Eq. \ref{eq:CR} dominates the impact of CR on the cluster cooling flows. All simulations included for CRs advection by the gas, CR pressure work, and CR radiative losses.

\begin{figure}
    \centering
    \includegraphics[width=\columnwidth]{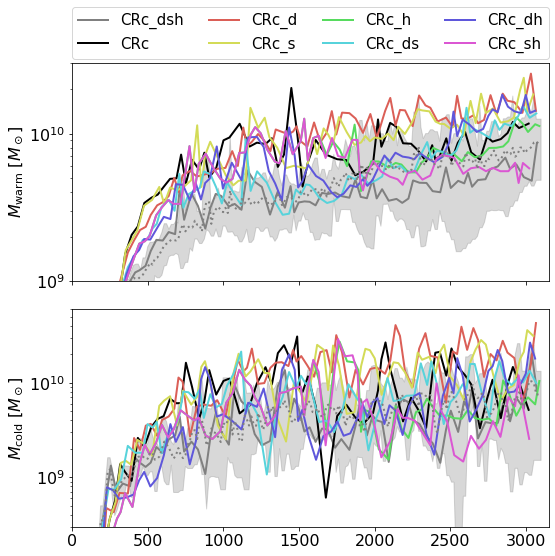}
    \caption{Time series for the total warm  ($10^4\,\rm K<T<10^6\,\rm K$) and cold ($T<10^4\,\rm K$) gas mass for different CR transport mechanisms in comparison to the range covered by all full-physics simulations based on CRc\_dsh presented in Appendix \ref{sec:stochastic} (grey area). The dotted grey line denotes the mean of the grey area at each point in time. Overall trends are difficult to differentiate when individual simulations are compared due to the chaotic nature of cooling flows.}
    \label{fig:transport_timeseries}
\end{figure}

Fig. \ref{fig:transport_timeseries} shows that different combinations of transport terms lead to variations in the evolution of the total cold and warm gas mass over time, but the chaotic nature of the simulations makes it difficult to isolate clear trends in comparison to the range of evolution covered by the full-physics simulations CRc\_dsh (filled grey area, based on all simulations from Appendix \ref{sec:stochastic}). The simulation with CRs but without any transport mechanisms (CRc) does not produce a strong cooling flow. Like for all simulations with only partial CR transport, $M_{\rm warm}$ is enhanced during the early evolution of the cluster, that is, while $t< 1.5 \rm \ Gyr$. However, the late-time evolution of CRc is consistent with the range of behaviour exhibited by CRc\_dsh for both warm and cold gas, without a sign of the run-away build-up of cold gas mass $M_{\rm cold}$ exhibited for example by MHDc (see Section \ref{fig:suite_gas_quantities}). 

This is in contrast with the results presented for a galaxy group by \citet{Wang2020}, where the case without streaming advection and streaming heating produced a strong run-away cooling flow that was not regulated by the AGN, which is only balanced when CR streaming was included. Being a galaxy group, rather than the massive galaxy cluster studied here, the hot gas in \citet{Wang2020} is cooler, which results in shorter cooling times, and the AGN jet is about an order of magnitude less powerful. We note that in the presence of CR streaming heating, the redistribution of CRs within the cluster due to CR streaming advection becomes inefficient, as can be seen by comparing CRc\_h and CRc\_sh in Fig. \ref{fig:transport_timeseries}, which only diverge after 1.5 Gyr of evolution. Therefore, the impact on cooling flows in \citet{Wang2020} is entirely due to streaming heating and not to a redistribution of CR by streaming advection. This suggests that CRs might play a very different role in galaxy groups and clusters of different masses. 

\begin{figure}
    \centering
    \includegraphics[width=\columnwidth]{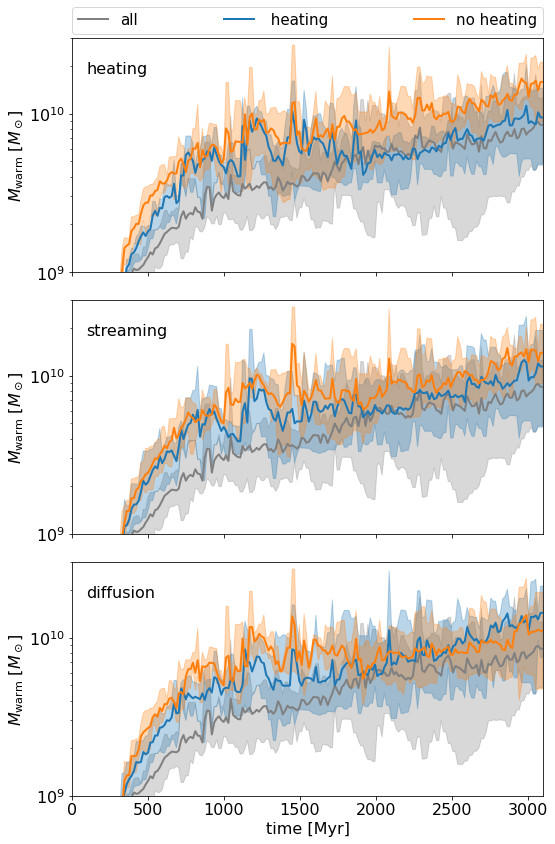}
    \caption{Average amount of warm gas $M_{\rm warm}$ over time for all simulations with (blue) in comparison to all simulations without (orange) a given CR transport mechanism. Average masses for full-physics simulations are shown in grey. Shaded areas denote the minimum to maximum $M_{\rm warm}$ at a given point in time for all simulations in a given bin. Only CR heating by the streaming instability has a significant impact on the evolution of the warm gas in the cluster, and only after $t> 1.5 \rm \ Gyr$.}
    \label{fig:transport_batched_timeseries}
\end{figure}

Despite the lack of cooling flows in CRc, CR transport mechanisms do influence the long-term evolution of the cluster, as can be seen in the binned time series in Fig. \ref{fig:transport_batched_timeseries}. When the average warm gas mass of all simulations that include the (streaming) heating term (top panel, blue line) are compared to all simulations that do not (top panel, orange line), it becomes clear that streaming heating dominates the late-time evolution of the warm gas. Beyond $t>1.5 \rm \ Gyr$, simulations with heating have the same amount of warm gas on average as those with all transport terms (grey), whereas simulation without streaming heating produce significantly more warm gas. Therefore, CR streaming heating is able to keep some of the gas hot that would otherwise cool into the warm phase. The CR radiative cooling times due to Coulomb and hadronic collisions ranges from 84 Myr ($n_{\rm e}=5 \times 10^{-1} \, \rm cm^{-3}$) to 4.2 Gyr ($n_{\rm e}=10^{-2} \, \rm cm^{-3}$) based on the cooling rate from \citet{Guo2008} in the hot gas, and can potentially be as short as 20 Myr in the warm gas ($n_{\rm e} = 2 \, \rm cm^{-3}$). For this reason, CR pressure is sustained over long periods of time, so that streaming heating provides a steady long-term source of energy that is able to maintain gas in the warm and hot phase for long periods of time. 
One notable feature of Fig. \ref{fig:transport_batched_timeseries} is that the average $M_{\rm warm}$ for full-physics simulations (grey line) continues to increase even after 3 Gyr of evolution. This suggests that the cluster has not yet reached a steady-state configuration. This raises an interesting question as to the eventual fate of this condensed gas. Baring a major reheating event, most of this gas will eventually have to cool to the cold phase, potentially producing a large amount of cold gas mass at later times. However, longer simulations runs by \citet{Ruszkowski2017} showed a continued gradual evolution of clusters under the influence of CRs on timescales of 6 Gyr or more, making a sudden cooling flow in CRc\_dsh unlikely.

Streaming advection itself has a weaker influence on cluster evolution. Simulations with the streaming advection term have half the amount of $M_{\rm warm}$ on average than those without, but still twice that of CRc\_dsh. Results that CR streaming and streaming heating play an important role in shaping long-term galaxy cluster cooling flows agree with those reported by \citet{Ruszkowski2017} and \citet{Wang2020}.

The last panel of Fig. \ref{fig:transport_batched_timeseries} shows that there is no significant difference in cooling flows with and without CR diffusion. The lack of importance of CR diffusion seems at odds with the conclusions from Section \ref{sec:phase}, which concluded that most of the gas is in a diffusion-dominated phase (i.e. $\chi>1$). The important difference are the length-scales considered in the two scenarios. Section \ref{sec:phase} is concerned with small-scale diffusion and its influence on locally thermally unstable regions of gas. By contrast, changing the large-scale cooling flow properties of the whole cluster would require diffusion on very large scales to redistribute CR in the cluster centre and flatten CR density profiles. Fig. \ref{fig:images_eta} shows that jets with an age of 200 Myr extend to around $50-70$ kpc. In comparison, the diffusion length over the same period of time is $L_{\rm D}=\sqrt{D_{\rm CR}t}= 8.14\,\rm kpc$, which is significantly smaller than the length of the jet. As a result, CR remain confined to a region in and around the jet cone (see Fig. \ref{fig:suite_images}), and the overall impact of diffusion on the long-term evolution of the cluster is greatly diminished. 

\begin{figure}
    \centering
    \includegraphics[width=\columnwidth]{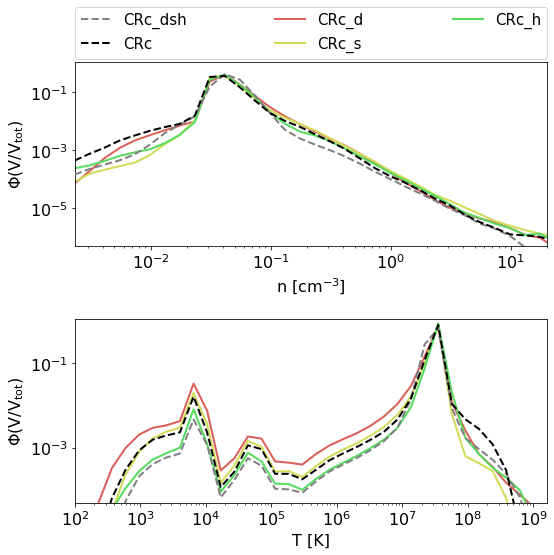}
    \caption{Volume-weighted probability distribution $\Phi(\rm V/V_{\rm tot})$ of the number density and temperature within the central 60 kpc of the cluster. Each line shows a stacked sample, measured at 0.25 Myr intervals throughout the simulation. While no CR transport mechanism influences the distribution of densities in the cluster centre, CR heating is required to reproduce the temperature distribution of the full-physics simulation.}
    \label{fig:transport_dens_T}
\end{figure}

The relative impact of the different CR transport terms is confirmed in Fig. \ref{fig:transport_dens_T}, which shows the distribution of gas densities (top panel) and temperatures (bottom panel) in the cluster centre. None of the transport terms has a significant impact on the distribution of gas densities, which is very similar for all stacked samples. However, with CR transport (CRc\_dsh), the distribution of temperatures is more peaked in the hot phase. The only simulation that can reproduce this trend is CRc\_h, that is, this peak in the temperature distribution is produced by streaming heating. A typical length scale on which heating deposits its energy can be found by comparing the streaming heating timescale $t_{\rm st}=(\gamma_{\rm CR}-1)L/u_{\rm A}$ with the diffusion timescale $t_{\rm D}=L^2/D_{\rm CR}$, which leads to $L=D_{\rm CR}/u_{\rm A}=974 (u_{\rm A}/\rm km\,s^{-1})^{-1}\,\rm kpc \approx 35 \rm \ kpc$  for our typical Alfvén velocity of $u_{\rm A} = 30 \rm \ km \ s^{-1}$. This value is approximately comparable with the extent of diffuse warm structures seen in Fig. \ref{fig:suite_images}. Increasing $D_{\rm CR}$ would increase both the efficiency of CR diffusion and the scale over which energy is deposited via streaming heating. 

Overall, we conclude that in our simulations, CR transport processes are not required to suppress cooling flows. Even simulations without these processes show the broad distribution of density and temperature and long-term self-regulation of cooling flows associated with a CR-dominated jet in our simulations. However, CR streaming heating, and to some extent, the redistribution of CR within the cluster centre through streaming advection, do influence the late-time evolution of the cluster cooling flow by helping to reduce both the cold and warm gas mass on timescales of 1.5 Gyr and longer.

\section{Thermal instability and cosmic rays}
\label{sec:phase}

In this section we study the relative importance of thermal, magnetic, and CR energy in the gas, and assess when and why criteria for thermal (in)stability are met in galaxy clusters for different fractions of CR energy injected in the jet, $f_{\rm CR}$. This section applies the thermal stability analysis of a thermal-CR composite fluid by \citet{Kempski2019}, which has been conducted only  in the linear regime. Recent work by \citet{Butsky2020} on the late non-linear evolution of thermal instability in the presence of CRs broadly confirmed results from \citet{Kempski2019}, but the full applicability of the analysis by \citet{Kempski2019} in the saturated non-linear regime found in the ICM in galaxy clusters is yet to be determined.

\begin{figure}
    \centering
    \includegraphics[width=0.8\columnwidth]{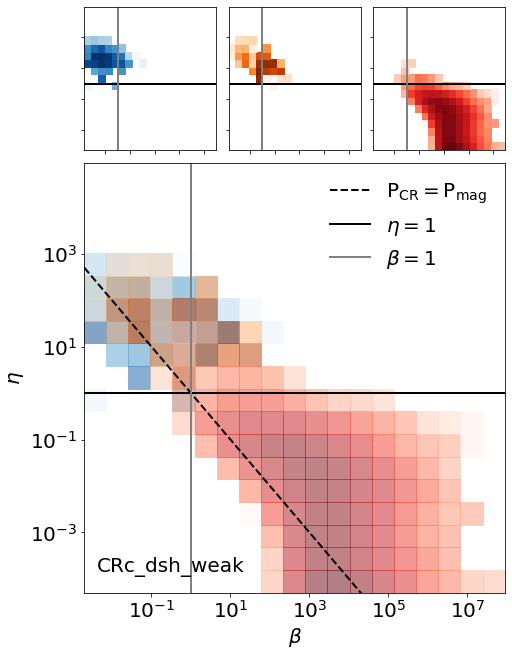} 
    \includegraphics[width=0.8\columnwidth]{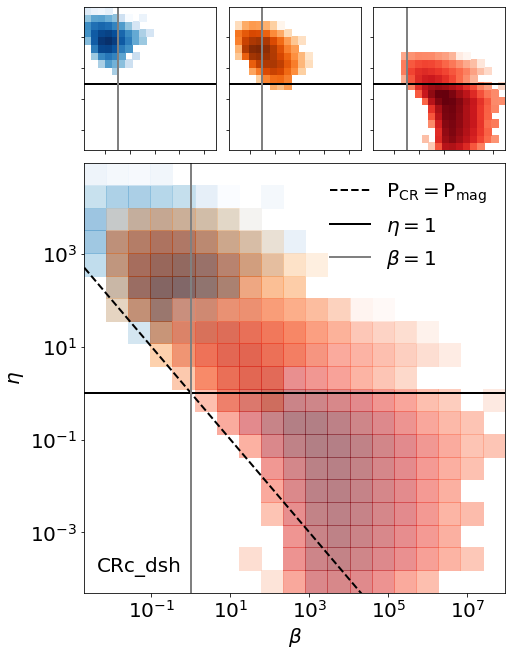}
    \caption{Mass-weighted distribution of plasma $\beta$ and CR pressure ratio $\eta$ for CRc\_dsh\_weak (top) and CRc\_dsh (bottom) for cold ($T<10^4 \rm \ K$, blue), warm ($10^4 < T < 10^6 \rm \ K$, orange), and hot ($T>10^6 \rm \ K$, red) gas, respectively. The small panels in the top row show only a single temperature bin, while all three temperature bins are superimposed in the large bottom panel. Above the dotted line, $P_{\rm CR} > P_{\rm mag} $ , while the reverse is true below. The data shown are for gas within the central 100 kpc of the cluster at time $t=2$ Gyr. All cold and warm gas is CR-pressure dominated, while all (most) of the hot gas is dominated by thermal pressure in CRc\_dsh\_weak (CRc\_dsh).}
    \label{fig:phase_eta_beta}
\end{figure}

Fig. \ref{fig:phase_eta_beta} shows that the main factor determining the relative importance of thermal pressure, CR pressure, and magnetic pressure is the underlying thermal phase of the gas: In general, thermal pressure dominates both magnetic pressure ($\beta>1$) and CR pressure ($\eta<1$) in the hot phase, but in the cold and warm phase, CR pressure dominates thermal pressure ($\eta>1$), and magnetic pressure contributes significantly or dominates  (i.e. $\beta<100$) in any fraction of CRs injected in the jet. As expected, a stronger injection of CR leads to a higher average CR fraction for cold and warm gas, and the wider spread in density and gas phases for the strong CR case also leads to a broader distribution of $\eta$ and $\beta$. However, when it is condensed out of the hot phase, all dense gas found here cools isochorically (i.e. has $\eta>1$), even for CRc\_dsh\_weak.

\begin{figure*}[t]
    \centering
    \begin{tabular}{cc}
    \includegraphics[width=0.5\textwidth]{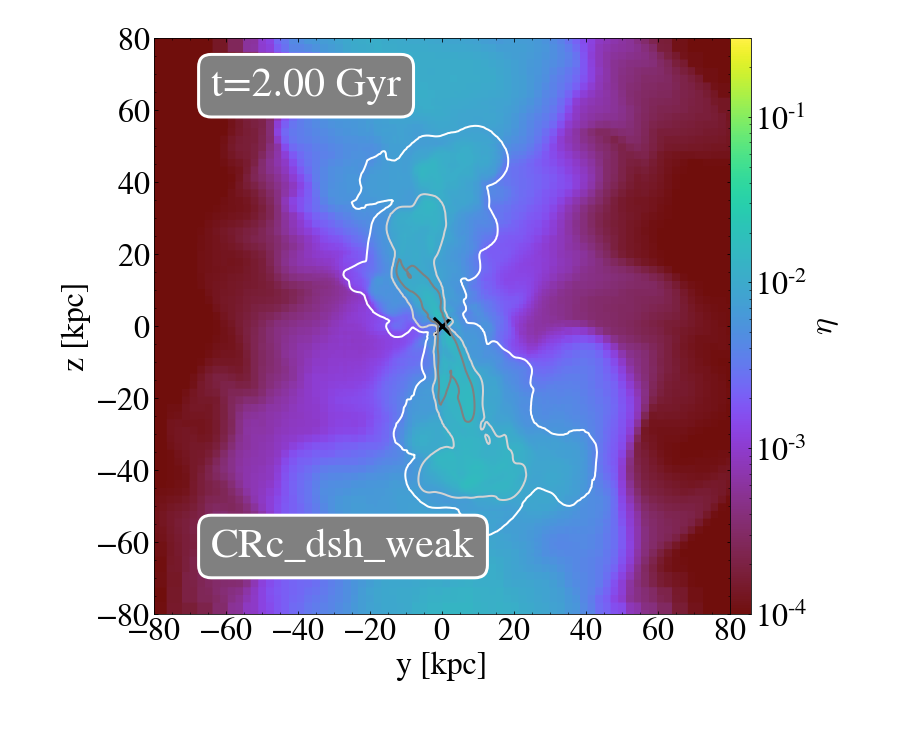} &
    \includegraphics[width=0.5\textwidth]{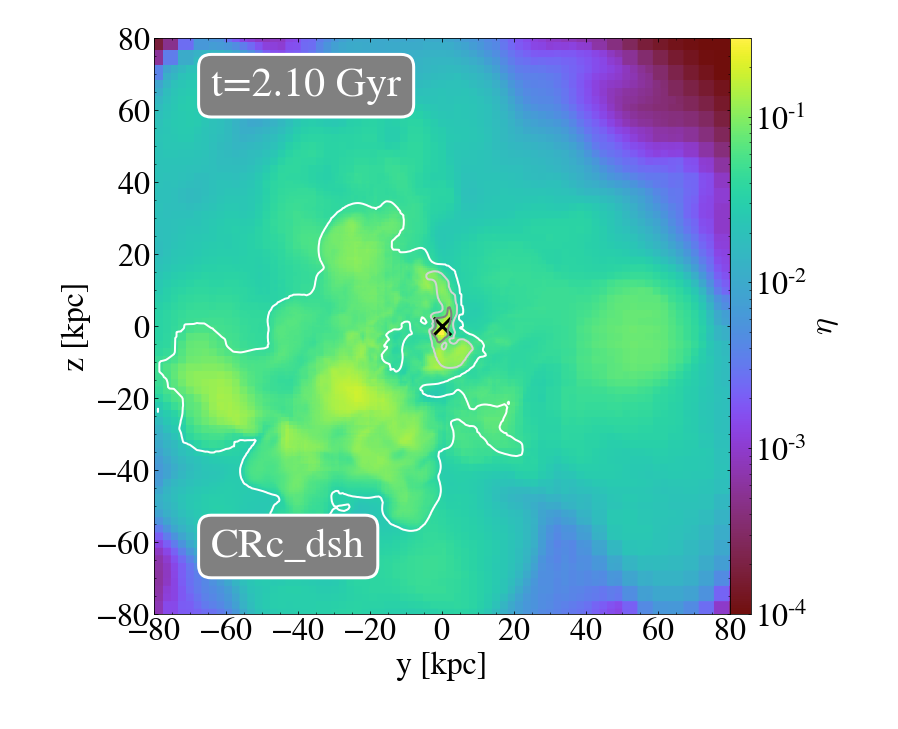} \\
    \end{tabular}
    \caption{Temperature-weighted projections of $\eta = {P_{\rm cr}}/{ P_{\rm therm}}$ for CRc\_dsh\_weak (left panel) and CRc\_dsh (right panel). Contours show the outline of remnants created by AGN jet ejections that occurred 200 Myr (white), 100 Myr (light grey), and 50 Myr (dark grey) before the snapshot. The black cross shows the current position of the AGN. The CR-dominated jet leads to an isotropic distribution of CRs in the cluster centre through the weak kinetic energy of the jet and the dense central nebula (right). For smaller fractions of CRs, the CRs remain confined to the jet cone (left).}
    \label{fig:images_eta}
\end{figure*}

Fig. \ref{fig:images_eta} shows that the spatial distribution of CRs in the hot gas is very different for the two cases. Because only 10\% of the energy is injected into CRs, the kinetic energy of the jet allows it to remain collimated and extend to much larger radii. As a result, CRs remain confined within a broad cylinder aligned with the jet axis, and $\eta$ is very low in regions far from the jet. 

By contrast, when 90\% of the jet energy is injected as CR energy, in CRc\_dsh, the volume-filling fraction of CRs is much higher. This is partially due to the lower kinetic energy of the jet. When 90\% of $L_{\rm AGN}$ is injected into CR energy, only 10\% are injected as kinetic energy, and CRc\_dsh generally already has a lower total $L_{\rm AGN}$ than CRc\_dsh\_weak (see Fig. \ref{fig:suite_AGN_timeseries}). With such high $f_{\rm cr}$ , jets carry much less momentum, which limits their ability to propagate to large distances from the cluster centre. Their propagation is further limited by the dense extended gas structure that fills the cluster centre out to 25 kpc or more (see Fig. \ref{fig:suite_images}). As a result, gas recently affected by AGN jets (white contours in Fig. \ref{fig:images_eta}) mixes more efficiently into the ICM, which leads to a more isotropic distribution of CRs in the cluster centre.

\begin{figure}
    \centering
    \includegraphics[width=\columnwidth]{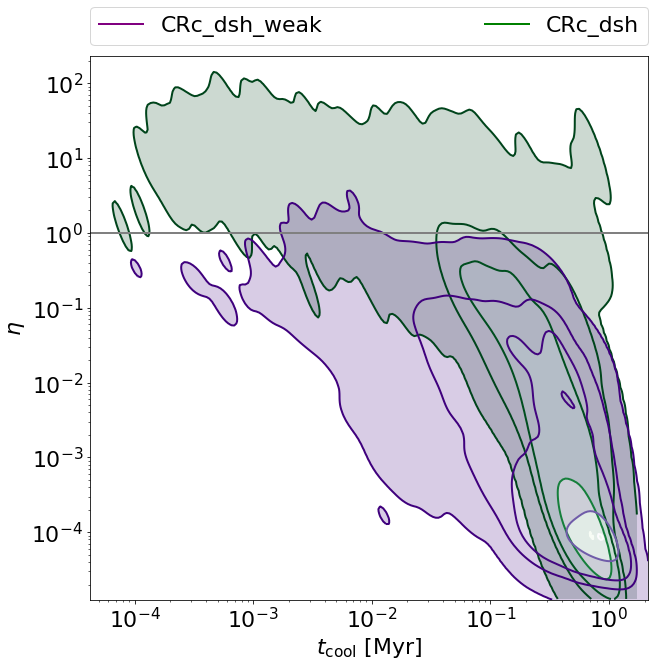}
    \caption{Volume-weighted kernel density estimate for cooling time $t_{\rm cool}$ vs $\eta$ for hot gas in CRc\_dsh\_weak and CRc\_dsh at time $t=2$ Gyr. Contours enclose 100\%, 99\%, 90\%, and 50\% of the total volume. Gas with the shortest cooling time has the highest ratio of CR to thermal pressure, and is therefore well saturated with CRs.}
    \label{fig:eta_tcool}
\end{figure}

One important impact of stronger CR injection is that a hot phase dominated by CR pressure develops, which is entirely absent in CRc\_dsh\_weak (see Fig.\ref{fig:phase_eta_beta}), and which will cool isochorically even before transitioning to the warm phase. The distribution of cooling time $t_{\rm cool}$ versus $\eta$ for the hot phase alone (Fig. \ref{fig:eta_tcool}) shows that this CR-dominated hot gas has a wide range of short cooling times,  but that the gas that is expected to cool and condense next (i.e. the gas with the shortest cooling time) also has the highest concentration of CRs. Therefore, newly cooled gas is well-saturated with CRs, which might explain the heating of filamentary nebulae in cluster centers \cite{Ruszkowski2018}. As gas cools, the thermal pressure drops and $\eta= {P_{\rm CR}}/{P_{\rm therm}}$ increases further, leading to the high values of $\eta$  in cold gas shown in Fig. \ref{fig:phase_eta_beta}.

\begin{figure}
    \centering
    \includegraphics[width=0.8\columnwidth]{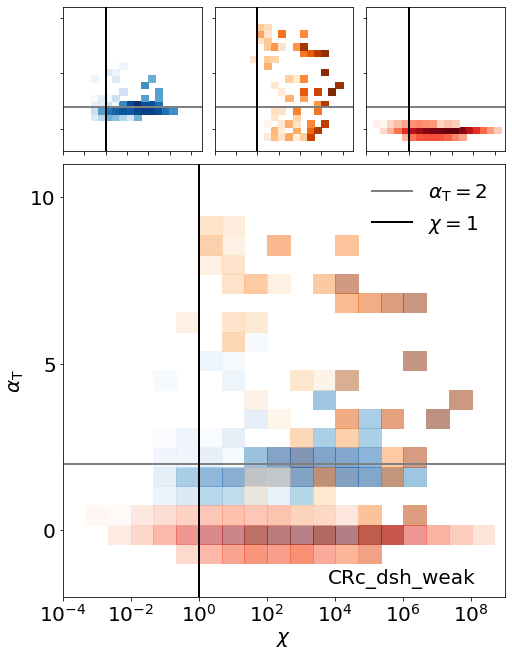} 
    \includegraphics[width=0.8\columnwidth]{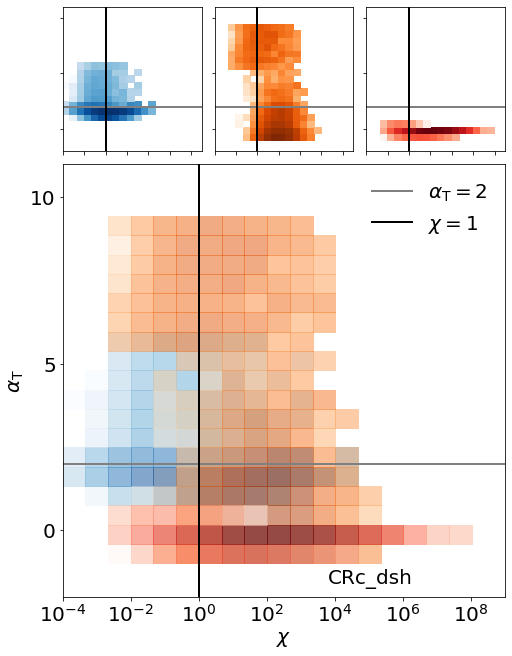} 
    \caption{Mass-weighted distribution of $\chi$ vs the slope of the cooling function $\alpha_{T}$ for cold ($T<10^4 \rm \ K$, blue), warm ($10^4 < T < 10^6 \rm \ K$, orange), and hot ($T>10^6 \rm \ K$, red) gas. The small panels in the top row show only a single temperature bin, while all three temperature bins are superimposed in the large bottom panel. The top plot shows values for CRc\_dsh\_weak, and the bottom panel shows them for CRc\_dsh.  Data are measured within the central 100 kpc of the cluster at time $t=2$ Gyr. As $\alpha_{\rm T} < 2$ for all hot gas, CRs are unable to prevent the onset of thermal instability in galaxy clusters.}
    \label{fig:CR_instability}
\end{figure}

As discussed in Section \ref{sec:theory}, the presence of magnetic fields enhances thermal instability, but the presence of CRs can stabilise small-scale thermal instability. Whether magnetised CR-saturated gas is thermally stable depends on the slope of the cooling function with respect to the temperature $T$, the power of the cooling function $\alpha_{T} = \partial\log(\mathcal{L}_{\rm th})/\partial\log(T)$, and the fraction $\chi = D_{\rm CR}  / (t_{\rm cool} \eta u_{\rm A}^2)$. CRs can suppress the onset of thermal instability if $\chi<1$, whereas thermally unstable gas remains so if $\chi>1$ and $\alpha_{T}<2$. 

As predicted, the bulk of the hot gas in both CRc\_dsh and CRc\_dsh\_weak can be found in the phase where CR make no difference to thermal instability ($\chi >1$, $\alpha < 2$, lower right quadrant, Fig. \ref{fig:CR_instability}). Only when the gas has already cooled to the warm phase, that is, when it has condensed out of the hot phase, does it reache $\alpha_{T}>2,$ where CR pressure can suppress instabilities and stabilise the gas against further collapse. CRc\_dsh\_weak has only a small amount of gas in the regime where $\chi <1,$ but in CRc\_dsh, a fraction of gas across all three phases (2\% of hot gas, 4\% of warm gas, and 38\% of cold gas, by mass) are found in this regime.  

\begin{figure*}[t]
    \centering
    \begin{tabular}{cc}
    \includegraphics[width=0.5\textwidth]{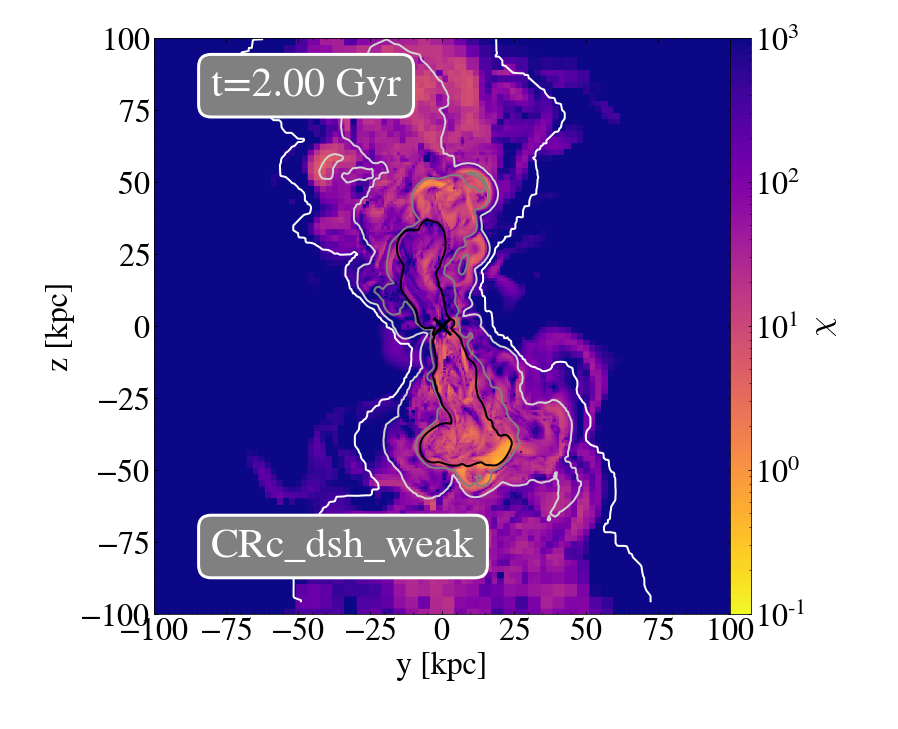} &
    \includegraphics[width=0.5\textwidth]{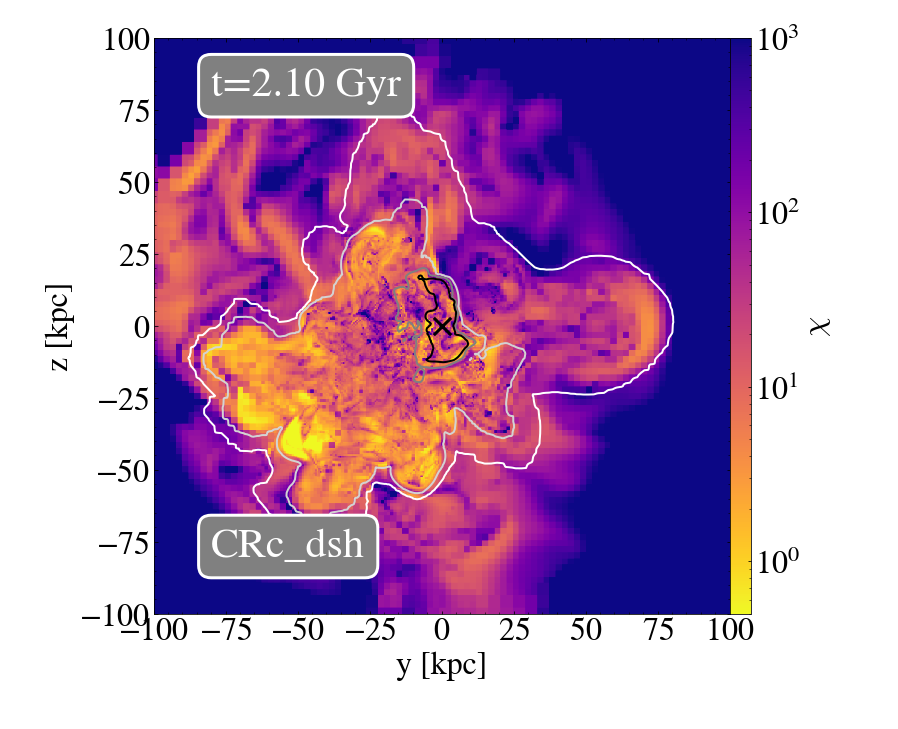} \\
    \end{tabular}
    \caption{Projections of $\chi$ weighted by the jet scalar (see Sec. \ref{sec:setup_AGN}) for CRc\_dsh\_weak (left panel) and CRc\_dsh (right panel). Contours show the outline of remnants created by AGN jet ejections that occurred 300 Myr (white), 200 Myr (light grey), 100 Myr (dark grey), and 50 Myr (black) before the snapshot. The black cross shows the current position of the AGN. The only diffusion-dominated gas (where $\chi<1$) is found within buoyantly rising AGN bubbles.}
    \label{fig:images_chi}
\end{figure*}

\begin{figure}
    \centering
    \includegraphics[width=\columnwidth]{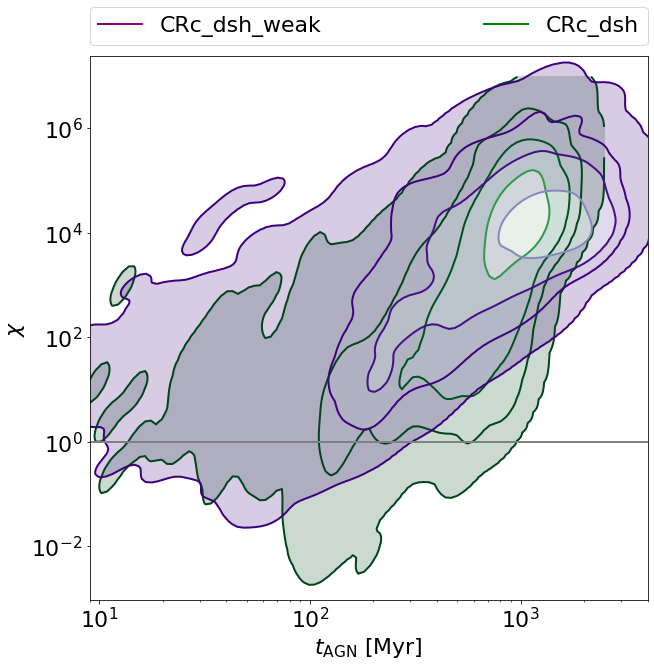}
    \caption{Volume-weighted kernel density estimated of  $\chi$ vs $t_{\rm AGN}$, the time since gas has been directly affected by AGN feedback for all hot gas within a sphere of radius 100 kpc, centred on the cluster centre. Contours enclose 100\%, 99\%, 90\%, and 50\% of the volume in the cluster centre. Gas most recently  affected by the AGN jet has the lowest $\chi$ values.}
    \label{fig:phase_chi_tAGN}
\end{figure}

This leaves the question where in the cluster this potentially stabilised gas is located, and whether stabilisation against thermal instability occurs on a sufficiently large scale to influence the cluster cooling flow. The projections in Fig. \ref{fig:images_chi} show that the lowest $\chi$ gas occurs within the AGN jet bubbles, in particular at their upper edges, where bubbles are rising buoyantly. This link between low $\chi$ and jet bubbles is shown more quantitatively in  Fig. \ref{fig:phase_chi_tAGN}, where the lowest $\chi$ values are found for the gas that has been most recently directly affected by the AGN jet (i.e. has low $t_{\rm AGN}$). Due to their long cooling times, driven by high temperatures and low densities, AGN jets and bubbles are locally thermally stable even in the absence of CRs \citep{Gaspari2011,Li2015,Bourne2019,Beckmann2019b}. Therefore, we conclude that the location of (thermal) instability within massive galaxy clusters is not meaningfully altered in the presence of CRs. Nonetheless, the thermodynamic evolution of gas after it has cooled out of the hot phase changes significantly with the number of CRs present in the gas. 

\begin{figure*}[t]
    \centering
    \includegraphics[width=\textwidth]{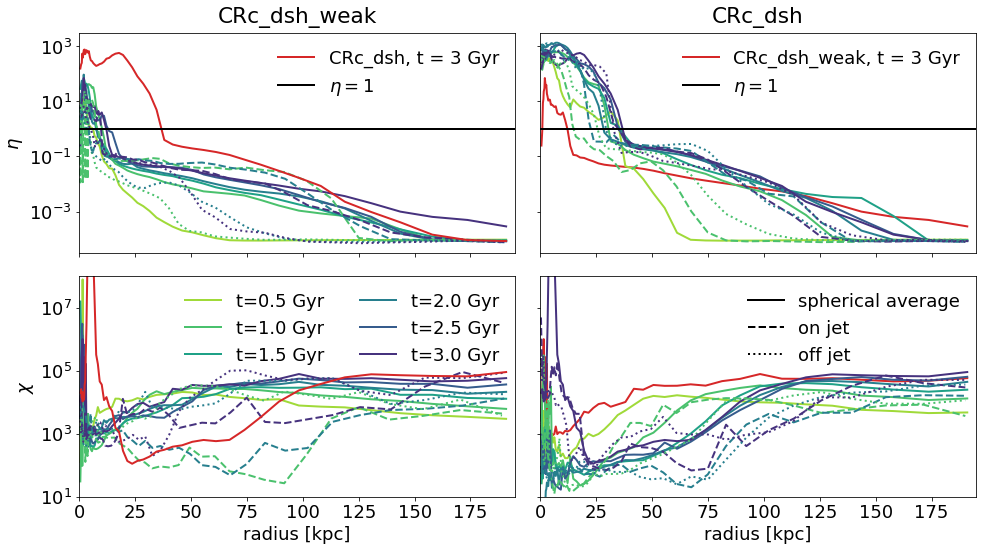}
    \caption{Time evolution of $\eta$ (top) and $\chi$ (bottom) profiles for CRc\_dsh\_weak (left) and CRc\_dsh. Solid profiles are a volume-weighted average over the whole sphere. At $t=1.5$ and 3 Gyr, we also show the profile measured along the jet (dashed) and along a cylinder perpendicular to the jet (dotted) for comparison. For comparison, the $t=3 $ Gyr value of the radial averaged profile for the other simulation is shown in red in each plot. In the CR-dominated and kinetic-dominated jets, the cluster core is dominated by CR pressure, but the extent of the CR-dominated core is much larger for a CR-dominated jet.}
    \label{fig:radial_eta_chi}
\end{figure*}

The radial distribution of $\eta$ and $\chi$ (Fig. \ref{fig:radial_eta_chi}, hot gas only) shows that the cluster core is dominated by CR pressure out to $\sim 10$ kpc for CRc\_dsh\_weak and out to $\sim 30$ kpc for CRc\_dsh. To quantify the variation in the two quantities depending on the location of the jet axis, we compared the average radial distribution for the whole sphere (solid line) to two cylinders with a diameter of 15kpc: "on-jet" is aligned with the z-axis of the simulation, which roughly aligns with the jet-axis of the simulation at this point in time (see the black contours in Fig. \ref{fig:images_chi} for both simulations), while "off-jet" is aligned with the y-axis of the simulation, and therefore oriented perpendicular to the jet axis. All measurements are volume weighted and were only performed for hot gas ($T>10^6$ K). 

Fig. \ref{fig:radial_eta_chi} shows that for both CRc\_dsh\_weak and CRc\_dsh the radial profile converges to a CR-dominated core ($\eta>1$) and gas-pressure-dominated cluster outskirts ($\eta<1$) after 1 Gyr of evolution.
The core is larger in CRc\_dsh, where it extends to 40 kpc, in comparison to CRc\_dsh\_weak, where it extends only to 15 kpc. In both simulations, $\eta= 10^{-4}$ at large radii due to the initial conditions of the simulation. The profiles in $\chi$ build up over time; the lowest values are found in the range $5<r<20 \rm \ kpc$ in CRc\_dsh\_weak and $5<r<60 \rm \ kpc$ in CRc\_dsh. In neither cluster is there a region where $\chi < 1$ on average. The fact that the radial profile of $\eta$ converges after $2-3$ Gyr agrees well with \citet{Ruszkowski2017}.

When we compare the radially averaged distribution (solid line) to the profiles along the jet (dashed) to perpendicular to the jet (dotted), the distribution for CRc\_dsh is almost isotropic. The distributions of $\eta$ and $\chi$ on-jet and off-jet are very similar to the radial average case at all three times we tested here. This is not the case for the jet-dominated CRc\_dsh\_weak case, where for $10 \rm \ kpc < r < 150 \rm \ kpc$, the off-jet region has a significantly lower $\eta,$ while $\chi$ of the jet is significantly higher than average. 

Overall, we conclude that the fraction of AGN luminosity injected into CRs has a significant and lasting impact on the properties of the multi-phase ICM. We report a long-term self-regulation of cooling flows, and convergence in cluster properties over timescales of billions of years for a jet-dominated configuration with only a small fraction of CR ($f_{\rm CR}=0.1$) and for a CR-dominated feedback mode ($f_{\rm CR} =0.9$). The former closely resembles the MHD-only case in morphology and distribution of dense gas, but the small contribution of CR pressure to total gas pressure prevents over-cooling and allows the cluster to remain in self-regulation for long timescales. Strong CR feedback leads to a strongly CR-dominated cluster core and to an isotropic volume-filling distribution of warm gas that slows cluster cooling down over long timescales. Using the linear stability analysis of \citet{Kempski2019}, we find no evidence for regions within the cluster in which CRs are able to prevent existing local thermal instability.

\section{Comparison to \gammaray~observations}
\label{sec:observation}

While it has been conclusively shown in this work that CRs can influence the long-term evolution of galaxy clusters, the question remains whether this evolution is supported by observational evidence. One way to observationally constrain the CR content of galaxy clusters is via the {\gammaray}s emitted by hadronic processes. 

To compare our cluster simulations to observations, we computed the expected hadronic \gammaray~photon flux \mbox{$ F_{\rm \gamma}=\int q_\gamma(\vec{r},E_\gamma)\,d E_\gamma\,d V$} from our simulations using the source function $q_\gamma$ from \citet{Pfrommer2004},
\begin{eqnarray}
\label{q gamma}
\lefteqn{
q_\gamma(\vec{r},E_\gamma)\simeq
\sigma_\mathrm{pp}\,c\,n_\mathrm{N}(\vec{r})\,\tilde{n}_{\rm CRp}(\vec{r})}  \\
  & & \times\,\frac{2^{4-\alpha_\gamma}}{3\,\alpha_\gamma} \,\left( \frac{m_{\pi^0}\,c^2}
      {\mbox{GeV}}\right)^{-\alpha_\gamma}
      \left[\left(\frac{2\, E_\gamma}{m_{\pi^0}\, c^2}\right)^{\delta_\gamma} +
      \left(\frac{2\, E_\gamma}{m_{\pi^0}\, c^2}\right)^{-\delta_\gamma}
      \right]^{-\alpha_\gamma /\delta_\gamma}\, , \nonumber
\end{eqnarray}
where $\sigma_{\rm pp}=32 (0.96+e^{4.4-2.4 \alpha_\gamma}) \rm \, mbarn$ is the effective cross section, $c$ is the speed of light, $n_{\rm N} = n_{\rm e}(r) / (1 - 0.5 X_{\rm He})$ is the target particle number density, $n_{\rm e}(r)$ is the electron number density taken from our simulation, $X_{\rm He}=0.24$ is the fraction of helium, $a_\gamma=4(\alpha_{\rm p}-0.5)/3$ is the $\gamma$-ray spectral index, $m_{\pi^0}$ is the pion mass, and $\delta_\gamma=0.14 \, \alpha_\gamma^{-1.6}+0.44$ is a shape parameter. 
\begin{equation}
\tilde{n}_{\rm CRp}(\vec{r}) = \varepsilon_{\rm CR}(\vec{r})\frac{2(\alpha_{\rm p}-1)}{ m_{\rm p}c^2} \left (\frac{m_{\rm p}c^2}{\rm GeV}\right)^{\alpha_{\rm p}-1} \mathcal{B}^{-1}\left(\frac{\alpha_{\rm p}-2}{2},\frac{3 - \alpha_{\rm p}}{2}\right)
\end{equation}
is an effective CR number density, where $\alpha_{\rm p}$ is the slope of the CR spectrum, $\varepsilon_{\rm CR}(\vec{r})$ is the kinetic CR energy density, taken from our simulation, $m_{\rm p}$ is the proton mass, and $\mathcal{B}(a,b)$ is the beta-function.

To compare to the observed $\gamma$-ray flux limits from \citet{Ackermann2014}, we integrated over the energy range $E_{\gamma} = [0.5,200] \rm \ GeV$. To calculate the flux, we assumed that our cluster is at the distance of the Perseus cluster, $73.8 \rm \ Mpc$. For the model cited here, $F_{\gamma}$ is at its maximum for $\alpha_{\rm p} = 2.2$, but $F_{\gamma} \rightarrow  0$ as $\alpha_{\rm p} \rightarrow 2$ and $\alpha_{\rm p} \rightarrow 3$. To establish upper limits, we adopted  $\alpha_{\rm p} = 2.2$ as our fiducial value, but also explored the impact of varying $\alpha_{\rm p}$.

\begin{figure}
    \centering
    \includegraphics[width=\columnwidth]{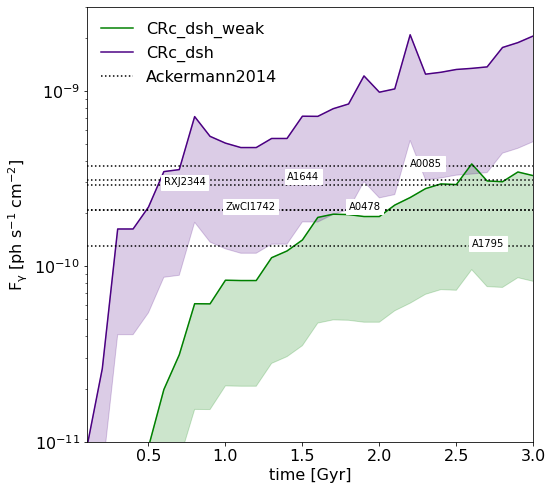}
    \caption{Time series of the total \gammaray~flux within 100 kpc from the cluster centre, $\rm F_{\gamma}$ , over time. Point-mass \gammaray~fluxes for all cool-core clusters with $6 \times 10^{14} \rm \, M_\odot < M_{200} < 10^{15} \rm \, M_\odot $ from \citet{Ackermann2014} are shown as vertical lines for comparison. A self-regulating CR-dominated jet produces a \gammaray~emission that exceeds observational limits, while small fractions of CR in the jet are compatible observational constraints.}
    \label{fig:gamma_timeseries}
\end{figure}

The resulting total \gammaray~energy flux is shown in Fig. \ref{fig:gamma_timeseries}. We also show for comparison point-mass \gammaray~fluxes for cool-core clusters with $6 \times 10^{14} \rm \, M_\odot < M_{200} < 10^{15} \rm \, M_\odot $ from \citet{Ackermann2014}.

Fig. \ref{fig:gamma_timeseries} shows that CRc\_dsh produces a significantly higher $F_\gamma$ than CRc\_dsh\_weak. Emission from CRc\_dsh\_weak is beginning to saturate from 2.5 Gyr onwards, whereas emission from CRc\_dsh continues to rise. When it is compared with the observed  upper limits for \gammaray~ fluxes from \citet{Ackermann2014}, CRc\_dsh quickly exceeds observational constraints by up to an order of magnitude. Emission from CRc\_dsh\_weak, by comparison, builds up much more slowly and converges at a value that agrees well with the observational upper limits.

We therefore conclude that for a self-regulated AGN, CR-dominated jets with high $f_{\rm CR}$ are not supported observationally for massive galaxy clusters. A small contribution of CR to the AGN luminosity of $f_{\rm CR} \leq 0.1,$ however, produces \gammaray~ emission that is compatible with current observational limits. This is true for a range of different slopes of the CR spectrum $2.03<\alpha_{\rm p}<2.73$ (shaded region, covering the range of $\alpha_{\rm p}$ such that $F_{\gamma} > 0.25 \ times F_{\rm \gamma, max}$). For CRc\_dsh to agree with the observed limits from \citet{Ackermann2014}, the CR spectrum would have to be either very shallow or very steep indeed.

In addition to heating and pressure support, CRs impact the gas chemistry, in particular in the cold molecular gas, where secondary electrons produced by CR dissociate CO molecules. Although we did not probe the molecular phase in these simulations, we anticipate that high $f_{\rm CR}$ values would help maintain a high C+/CO abundance \citep{Mashian2013}. This could explain the very strong C+ cooling lines observed in brightest cluster galaxies \citep[e.g.][]{Edge2010, Mittal2012, Polles2021}.  

\section{Discussion and conclusions}
\label{sec:conclusions}

We studied how injection of a fraction of the luminosity of the central AGN of a galaxy cluster as CRs influences the cooling flow of a massive galaxy cluster. We focused on how changing the fraction of AGN jet energy that is injected into CR energy rather than in kinetic energy, $f_{\rm CR}$, changes the evolution of local thermal instability on timescales of billions of years, and what impact the resulting CR pressure and heating have on the multi-phase structure of the ICM. Our results are listed below.

\begin{enumerate}
    \item Generally, injecting a fraction of the AGN luminosity as CR energy at the jet base aids the AGN in regulating cooling flows on timescales of billions of years. Even values as low as $f_{\rm CR} \leq 0.1$, that is, a case in which only 10\% of the AGN luminosity is injected as CRs, prevent thermal instability and strong cooling flows.  
    \item For a CR-dominated jet (here $f_{\rm CR}=0.9,$ i.e. 90\% of $L_{\rm AGN}$ is injected as CR energy), the structure of multi-phase gas changes significantly. Rather than cooling very quickly from the hot, diffuse phase to a cold, dense phase, gas spends long periods of time in a warm, diffuse phase and remains at lower densities even at cold temperatures. This is due to the additional pressure support from CRs and to the energy transfer from CRs to thermal energy via streaming instability heating. 
    \item Within massive galaxy clusters, CRs are distributed mainly via advection and not via diffusion or streaming. For low $f_{\rm CR}$, CRs are concentrated in and around the AGN jet. For high $f_{\rm CR}$, the low kinetic energy in combination with the build-up of condensed gas in the cluster centre means that jets struggle to propagate to large distances and deposit their energy and CRs almost isotropically within the central 70 kpc of the cluster.
    \item With any $f_{\rm CR}$ tested here, hot gas in the cluster remains dominated by thermal pressure over CR pressure and magnetic pressure. Conversely, cold and warm gas is dominated by CR pressure for any $f_{\rm CR}$.
    \item Cosmic rays are unable to suppress local thermal instability in the hot gas because the conditions for them to do so, according to the linear perturbation analysis derived in \citet{Kempski2019}, are only met within the buoyantly rising outer edges of the AGN bubbles, which are thermally stable already due to their long cooling times.
    \item Self-regulating CR-dominated jets (i.e. those that have high $f_{\rm CR}$) lead to \gammaray~ emission beyond current observational upper limits. Values as low as $f_{\rm CR}=0.1,$ however, are still within observational upper limits, and therefore represent the more likely scenario for massive galaxy clusters.
\end{enumerate}

A low fraction of CRs in the jet also agrees well with predictions from particle-in-cell simulations \citep[e.g.][]{Crumley2019}, which show that only about 10 \% of the kinetic energy can be converted into CRs in strong shocks. Higher CR fractions can only be explained if at injection scales, $f_{\rm CR}<0.1$, but that by the resolution scale of the simulation (here $\Delta x \sim 500$ pc) sufficient kinetic energy has been transferred to gas internal energy and radiated away such that the remaining energy balance in the jet has shifted significantly. For strongly cooling gas, this scenario is certainly possible, but it does imply that the jet was significantly more powerful at injection. For example, for a jet to change from 20\% CRs to 90\%, it must have been 4.5 times as powerful at injection and must have lost 77\% of its original energy by 500 pc. 

Overall, we conclude that if up to 10\% of the AGN luminosity is injected as CR energy, cooling flows self-regulate over timescales of billions of years due to the additional pressure support and to the additional heating of the gas via CR streaming for gas that condenses out of the hot phase. In this scenario, multi-phase gas properties look globally similar to the extensively studied non-magnetised, CR-free case, without evidence for extensive diffuse, warm gas seen for gas with higher CR pressures. 

\begin{acknowledgements}
RB And YD designed the project. RB, YD and AP contributed to code development. RB designed, executed and processed the suite of simulations. RB and YD analysed and interpreted results, and wrote the paper. VO, FP, ML, OH and PG provided observational context and discussion of results. We would like to thank the anonymous referee for their comments.\\

This work was supported by the ANR grant LYRICS (ANR-16-CE31-0011) and was granted access to the HPC resources of CINES under the allocations A0080406955 and A0100406955 made by GENCI. RB would like to thank Newnham College, University of Cambridge, for support. AP acknowledges funding from the European Research Council (ERC) under the European Union’s Horizon 2020 research and innovation programme (grant agreement No. 679145, project ‘COSMO-SIMS’). This work has made use of the Infinity Cluster hosted by Institut d’Astrophysique de Paris. We thank Stéphane Rouberol for smoothly running this cluster for us. Visualisations in this paper were produced using the yt project \citep{Turk2011}.  
\end{acknowledgements}




\bibliographystyle{aa}
\bibliography{author} 




\begin{appendix}
\section{Cooling function}
\label{sec:cooling_function}

\begin{figure}
    \centering
    \includegraphics[width=0.9\columnwidth]{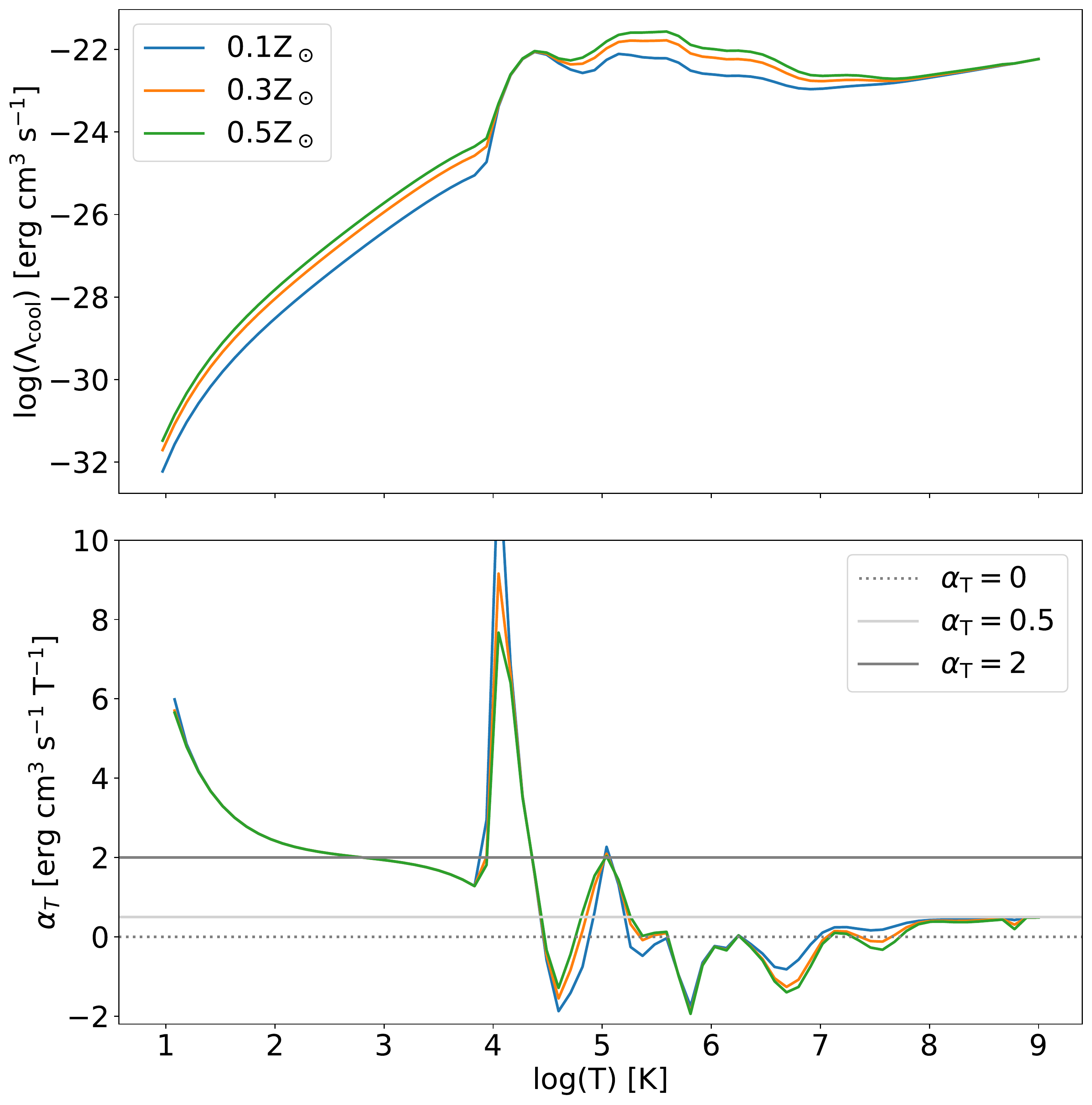}
    \caption{Cooling function $\mathcal{L}_{\rm cool}$ (top panel) and $\alpha_T$ (bottom panel) used in the simulations presented here, as a function of temperature $T$ and metallicity $Z$.}
    \label{fig:cooling_function}
\end{figure}

 Fig. \ref{fig:cooling_function} shows that the cooling function used in the simulations presented here, based on \citet{Sutherland1993} and \citet{Rosen1995}, is a function of temperature and metallicity, but is entirely independent of number density for $10^{-2} < \rm n < 10^{2} \rm \ cm^{-3}$. As a consequence, the same is true for its slope $\alpha_{\rm T}$. In general, for massive galaxy clusters, $\alpha_{\rm T}>2$ only around temperatures of $10^{4}$ K.

\section{Stochastic nature and robustness of results}
\label{sec:stochastic}

Cluster cooling cycles are inherently chaotic. 
The energy input via the AGN jet is driven not by the overall amount of cold gas, but by the consistent supply of cold gas very close to the BH. The zone of influence of the BH is small (at the resolution limit for our simulations). The total injected energy therefore not only depends on global cooling, but also, among others, on the distribution of cold gas within the cluster centre and on the position of the BH. In this section, we test how robust the trends observed in the previous sections are to chaotic perturbations of the system over long periods of time. 

To do this, we conducted five variations of simulation CRc\_dsh, the details of which can be found in Table \ref{tab:AGN_simulations}:
\begin{itemize}
    \item CRc\_dsh\_1 and  CRc\_dsh\_2 have identical parameter choices to CRc\_dsh, but initial conditions were set up with a different random seed for the initial tangled magnetic field and velocity perturbations.
    \item CRc\_dsh\_f0.5 and CRc\_dsh\_f1 have identical initial conditions to CRc\_dsh, but have a higher pre-factor for the gradient descent acceleration. In these simulations, the BH is anchored more securely to the cluster centre.
    \item CRc\_dsh\_f1\_1 is identical to CRc\_dsh\_f1 except that sub-grid drag forces due to gas and particles have been switched off to test their impact on the dynamics of the BH.
\end{itemize}

\begin{figure}
    \centering
    \includegraphics[width=0.9\columnwidth]{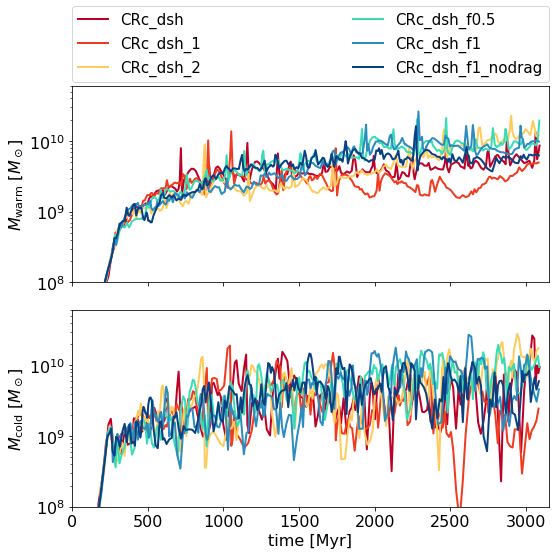}
    \caption{Time evolution of the total warm $10^4 < T < 10^6 \rm \ K$ and cold $T<10^4 \rm \ K$ gas mass for variations of the simulations CRc\_dsh. The time evolution of the cold and warm gas mass is robust to perturbations in AGN dynamics.}
    \label{fig:stochastic_cooling_flow}
\end{figure}

\begin{figure}
    \centering
    \includegraphics[width=0.9\columnwidth]{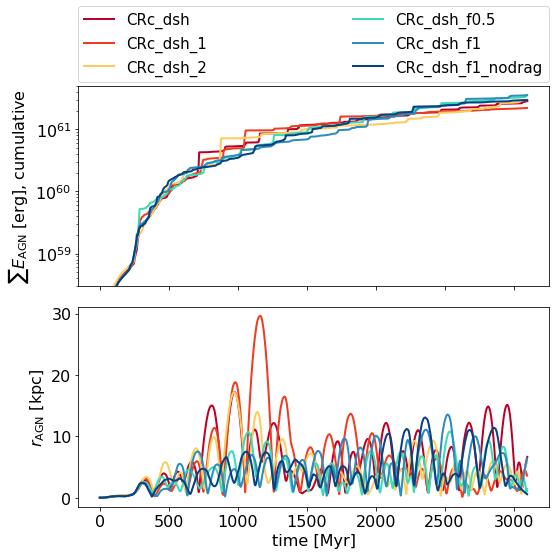}
    \caption{Time series of cumulative AGN energy injected throughout the simulation and distance of the AGN from the cluster centre for variations of the simulations CRc\_dsh. The overall energy input required for cluster self-regulation is very similar even for different AGN dynamics. }
    \label{fig:stochastic_AGN_timeseries}
\end{figure}

Fig. \ref{fig:stochastic_cooling_flow} confirms that the global evolution of gas phases is similar for all six simulations. While the turbulent variations in the warm (top panel) and cold (bottom panel) gas mass is readily apparent when the simulations are compared, no overall trends or strong divergence of results emerge over the 3 Gyr of evolution probed here. This conclusion is confirmed by the AGN time series in Fig. \ref{fig:stochastic_AGN_timeseries}.
The AGN luminosity, picking up the variations in mass and distribution of the condensed gas, varies strongly but remains within the same overall luminosity range for all six simulations. The overall AGN energy injected into all six simulations after 3 Gyr of simulation is even more robust. It ranges from $2.2 \times 10^{61} \rm \, erg\,s^{-1}$ for CRc\_dsh\_1 to $3.5 \times 10^{61} \rm \, erg\,s^{-1}$ for CRc\_dsh\_f1 in Fig. \ref{fig:stochastic_AGN_timeseries}. We therefore conclude that overall, the conclusions drawn in this paper are robust to reasonable perturbations of the system, despite its overall chaotic nature.

\section{Cosmic-ray parameters}
\label{sec:varCR}

\begin{figure}
    \centering
    \includegraphics[width=0.9\columnwidth]{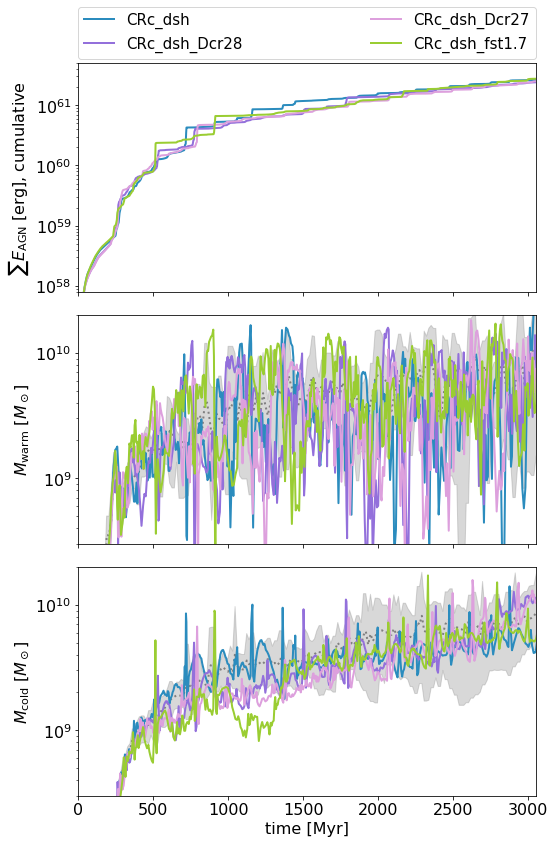}
    \caption{Time evolution of the cumulative energy injected by the AGN (top) and the warm (middle) and cold (bottom) gas mass for the three variations of CRc\_dsh listed in Table \ref{tab:varCR}. Grey shaded areas show the statistical variation of CRc\_dsh from Appendix \ref{sec:stochastic}. Neither a lower $D_{\rm CR}$ nor a higher $f_{\rm b,st}$ change the cooling flow properties of the clusters beyond stochastic variation.}
    \label{fig:varC_evolution}
\end{figure}

\begin{table}[!b]
    \centering
    \begin{tabular}{lcc}
         \textbf{Simulation name} & $\log(D_{\rm CR})$ [$\rm cm^{-2} s^{-1}$]  & $f_{\rm b,st}$ \\
         \hline
         CRc\_dsh$^*$ & $29$ & 1 \\
         CRc\_dsh\_Dcr28 & $28$ & 1 \\
         CRc\_dsh\_Dcr27 & $27$ & 1 \\
         CRc\_dsh\_fst1.7 & $29$ & 1.7 \\
    \end{tabular}
    \vspace{0.3cm}
    \caption{Variation of CRc\_dsh to explore the parameter choices of the CR model. All parameters not listed in this table are identical to CRc\_dsh. The simulation with the asterisk is included as a reminder.}
    \label{tab:varCR}
\end{table}

To ensure that our results do not depend sensitively on the parameters chosen as part of the CR model, in this section we conduct a series of variations of our fiducial simulation, CRc\_dsh, with different values for $D_{\rm CR}$ and $f_{\rm b,st}$ (see Sec. \ref{sec:CRphysics}), to determine how sensitively our results depend on these two parameters. The CR diffusion coefficient $D_{\rm CR}$ is poorly constrained for astrophysical plasma because it sensitively depends on the small-scale morphology of the magnetic field on length scales of the order of the Larmor radius. This is an estimated $10^8$ orders of magnitude below the resolution of our simulation (\cite{Snodin2016}). We therefore relied on a simplified description of diffusion, whose efficiency is captured in the magnitude of $D_{\rm CR}$. For our fiducial simulation, we selected $D_{\rm CR} = 10^{29} \rm \ cm^{-2} s^{-1}$, based on recent work by \cite{Chan2019} (see also \cite{Hopkins2021}), to constrain $D_{\rm CR}$ through a comparison of simulations to observations. A second free parameter is $f_{\rm b,st}$, which depends on the efficiency of damping processes that are excited by the streaming instability. Landau damping is thought to be the dominant mechanism for typical ICM properties, which allows for (moderate) super-Alfvénic streaming speeds such that $f_{\rm b,st}=1.7 $ (\cite{Ruszkowski2017}). To test the robustness of our results with respect to these two parameters, we examined three variations of our fiducial simulation: lower diffusion coefficients of  $D_{\rm CR} = 10^{27} \rm \ cm^{-2} s^{-1}$ (CRc\_dsh\_Dcr1E28) and  $D_{\rm CR} = 10^{28} \rm \ cm^{-2} s^{-1}$ (CRc\_dsh\_Dcr1E27), and an increased streaming boost factor of $f_{\rm b,st}=1.7$ (CRc\_dsh\_fst1.7). The parameter choices are summarised in Tab. \ref{tab:varCR}.

Fig. \ref{fig:varC_evolution} shows that lowering $D_{\rm CR}$ by up to two orders of magnitude or increasing $f_{\rm b,st}$ to $f_{\rm b,st}=1.7$ has little effect on the cooling flow in the cluster. The total energies injected by the AGN in each case are very similar, and the time series of $M_{\rm warm}$ and $M_{\rm cold}$ remain within the range for the stochastic variation of CRc\_dsh shown in Appendix \ref{sec:stochastic}.

\section{Resolution study}
\label{sec:resolution}

\begin{figure}
    \centering
    \includegraphics[width=0.9\columnwidth]{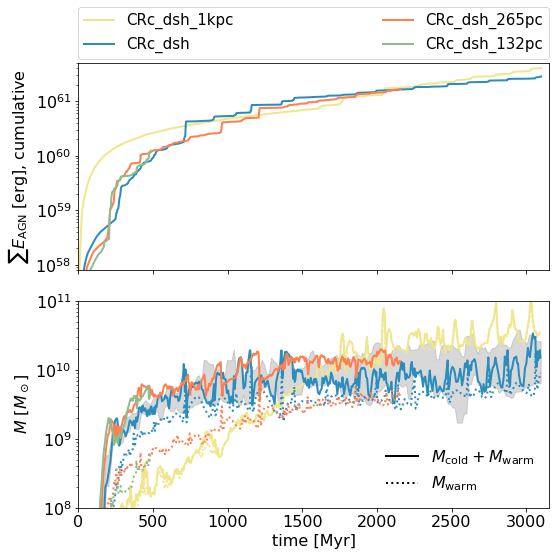}
    \caption{Time evolution of the cumulative energy injected by the AGN (top) and the warm and cold (solid bottom) and warm gas only (dotted bottom) gas mass over time. Grey shaded areas show the statistical variation of CRc\_dsh from Appendix \ref{sec:stochastic}. The case with the lowest resolution,  CRc\_dsh\_1kpc, requires significantly more energy to balance cooling flows at both early and late times, and produces predominantly warm gas. All three runs with higher resolution show good convergence in the cumulative AGN energy and total cold and warm gas. A higher resolution leads to more cold gas vs warm gas early on.}
    \label{fig:res_evolution}
\end{figure}

\begin{figure*}[btp]
    \centering
    \includegraphics[width=\textwidth]{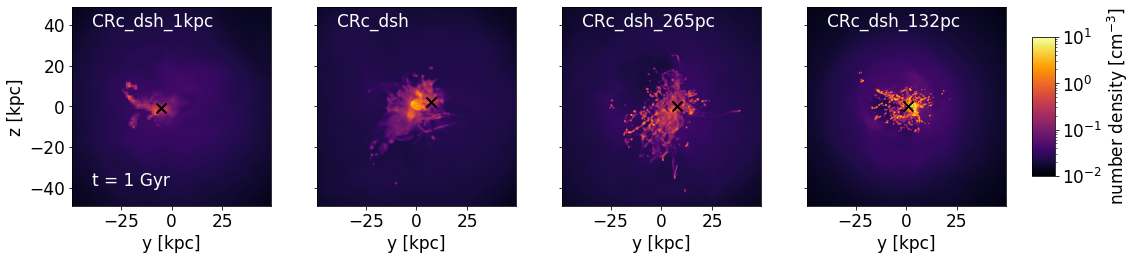}
    \caption{Projected gas density showing how the multi-phase structure of the gas changes with resolution.}
    \label{fig:res_images}
\end{figure*}

\begin{figure*}[btp]
    \centering
    \includegraphics[width=\textwidth]{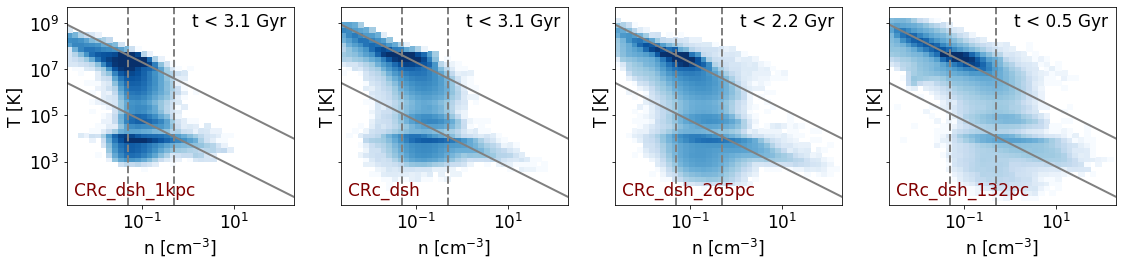}
    \caption{2D phase plots showing the volume-weighted distribution of gas number density $n$ vs temperature $T$ within the central 60 kpc of the cluster for a stacked sample measured every 0.25 Myr between $t=0$ Gyr and $t=3$ Gyr. Solid (dashed) grey lines show example lines along which gas evolves when it cools in an isobaric (isochoric) fashion. Histograms along the x-axis (y-axis) show 1D histograms for the number density (temperature), weighted by $V/V_{\rm tot}$ , where $V_{\rm tot}$ is the total volume probed here. Solid lines show the distribution at individual points in time, and the dashed line shows the distribution for the stacked sample. With increasing resolution, the distribution becomes broader, but the generally isochoric nature of cooling remains.}
    \label{fig:res_phaseplot}
\end{figure*}

To test whether our results are strongly dependent on resolution, we present three variations of our fiducial simulation CRc\_dsh with different maximum resolutions $\Delta x$. As a reminder, CRc\_dsh has $\Delta x= 531 \rm \ pc$ (refinement level 14). In this section, we also present CR\_dsh\_1kpc, with $\Delta x = 1.06 \rm \ kpc$ (level 13), CR\_dsh\_265pc ($\Delta x = 265 \rm \ pc$, level 15), and CRC\_dsh\_132pc ($Delta x = 132 \rm \ pc$, level 16). 

Fig. \ref{fig:res_evolution} shows that the results converge reasonably well for a simulation of this complexity, but only just. CRc\_dsh\_1kpc, the run with a lower resolution than the fiducial simulation, shows that a signficiant increase in cumulative AGN energy is needed to balance the cooling flow at early ($t<1 \rm \ Gyr$) and late ($t>1 \rm \ Gyr$) times. This simulation produces very little cold gas overall (see the difference between the dashed and dotted line in the bottom panel of Fig. \ref{fig:res_evolution}), and  while the onset of warm gas production is delayed in comparison to CRc\_dsh, it does not seem find a stable configuration, with $M_{\rm warm} + M_{\rm cold}$ continuing to increase after 3 Gyr. 

By comparison, the two simulations with a higher resolution, CRc\_dsh\_265pc and CRc\_dsh\_132pc, show a good convergence. The cumulative AGN energy required for all three is very similar, as is $M_{\rm warm} + M_{\rm cold}$ within the bounds of the expected stochastic variation of the cluster. The only notable trend is a reduction in $M_{\rm warm}$ and a corresponding increase in $M_{\rm cold}$ with increasing resolution early on. 

This difference in the distribution of gas is also visible in the density projection in Fig. \ref{fig:res_images}. A lower resolution leads to a more diffuse central nebula that is less extended at early times. Increasing the resolution allows for more clumping of the gas, which in turn means a broader distribution in temperature - density space, as is shown in Fig. \ref{fig:res_phaseplot}. However, the general evolution of cooling, that is, isobaric at high and low temperatures, but a broad isochoric distribution in between, is captured for the whole range of resolutions we tested here.

We therefore conclude that for $\Delta x < 500 \rm \ pc$, large-scale cooling flows in our galaxy clusters are well resolved. The internal structure of the multi-phase gas continues to evolve, however. 

\end{appendix}


\end{document}